\documentclass[12pt]{article}
\usepackage{latexsym}
\usepackage{epsfig}

\usepackage{graphicx}
\usepackage{epstopdf}

\usepackage{epsfig,amssymb,amsmath,amscd}

\newcommand{\bq}{\begin{equation}}
\newcommand{\eq}{\end{equation}}
\newcommand{\bqa}{\begin{eqnarray}}
\newcommand{\eqa}{\end{eqnarray}}
\newcommand{\ben}{\begin{enumerate}}
\newcommand{\een}{\end{enumerate}}
\newcommand{\bc}{\begin{center}}
\newcommand{\ec}{\end{center}}
\newcommand{\bqb}{\begin{eqnarray*}}
\newcommand{\eqb}{\end{eqnarray*}}

\def\pr#1#2#3{Phys. Rev. ${\bf{#1}}$, #2 (#3)}
\def\prl#1#2#3{Phys. Rev. Lett. ${\bf{#1}}$, #2 (#3)}
\def\pl#1#2#3{Phys. Lett. ${\bf{#1}}$, #2 (#3)}

\def\np#1#2#3{Nucl. Phys. ${\bf{#1}}$, #2 (#3)}

\def\jmp#1#2#3{J. Mod. Phys. ${\bf{#1}}$, #2 (#3)}

\begin{document}
\pagenumbering{arabic}
\thispagestyle{empty}
\def\thefootnote{\fnsymbol{footnote}}
\setcounter{footnote}{1}

\begin{flushright}
Feb.28, 2017\\
 \end{flushright}

\begin{center}
{\Large {\bf CSM constraints on the $gg\to ZH$ process}}.\\
 \vspace{1cm}
{\large F.M. Renard}\\
\vspace{0.2cm}
Laboratoire Univers et Particules de Montpellier,
UMR 5299\\
Universit\'{e} Montpellier II, Place Eug\`{e}ne Bataillon CC072\\
 F-34095 Montpellier Cedex 5, France.\\
\end{center}

\vspace*{1.cm}
\begin{center}
{\bf Abstract}
\end{center}

We study more deeply the sensitivity of the $ZH$ production process
in gluon-gluon collisions to the details of a possible Higgs and top
quark compositeness.
We establish a relation between Higgs and top quark form factors which would
keep the basic cancellation appearing in SM and satisfy the
concept of CSM. We give illustrations showing the spectacular consequences of
various choices preserving or violating this CSM constraint.

\vspace{0.5cm}

\def\thefootnote{\arabic{footnote}}
\setcounter{footnote}{0}
\clearpage

\section{INTRODUCTION}

In a previous paper \cite{ggZH} we have shown that the process of $ZH$ production
appearing at one loop in gluon-gluon or in photon-photon collision
is particularly sensitive to compositeness of the Higgs sector.\\
Various aspects of compositeness (in the Higgs and in other sectors)
can be found in ref.\cite{Hcomp, comp, Hcomp2, Hcomp3, partialcomp, Hcomp4}.\\
But without following one of these models we had assumed the possibility of
a compositeness concept (that we had called CSM in \cite{CSM}) which should
preserve the main features of the Standard Model (SM), at least at low energy.
No anomalous coupling modifying the SM structures should be present.  
Illustrations were given in which Higgs and Goldstone
couplings ($G^0Z_LH$ and $ZZ_LH$) are simply affected by compositeness
form factors.
They showed that the process $gg\to ZH$ is
especially sensitive to any small modification of the SM amplitudes because
that would destroy the peculiar cancellations between one loop triangle
and box diagrams of Fig.1. Indeed the introduction of $G^0Z_LH$ and $ZZ_LH$
form factors immediately destroys this specific cancellation and leads
to a strong enhancement of the rate of the  $gg\to ZH$ process.\\
Such cancellations between contributions of different sectors are not
exceptional in SM. The most famous case is $V_LV_L \to V_LV_L$ scattering
which would violate unitarity when the Higgs boson contribution
is not included \cite{VVVV}.\\

In the present paper we analyze this feature of the $gg\to ZH$ process in 
more details, i.e. the role
of each diagram and of its specific couplings. We extend the possibility of
compositeness to the top quark sector and we establish a relation between
form factors that would preserve the essential cancellation and could be
specific of a CSM picture.\\
We study the effects of this CSM constraint in various situations
with either only $t_R$ compositeness or with both $t_L$ and $t_R$ compositeness
and with typical choices
of form factors in the Higgs and top quark sectors.\\
We give illustrations showing the energy and angular dependences
of the various amplitudes and cross sections for polarized or unpolarized
$Z$ production.\\
Indeed spectacular differences appear between the CSM conserving and the CSM
violating cases.\\

Contents: In Section 2 we analyze in details the origin of the behaviours
of the various SM amplitudes and the consequences for the energy and angular
dependences of the cross section and of the rate of $Z_L$ production.
In Section 3 we introduce various form factors and establish the CSM
constraint. Illustrations comparing the effects of these various choices
are given in Section 4 and conclusions are summarized in Section 5.\\

\section{CHARACTERISTICS OF THE $gg\to ZH$ PROCESS}

It is important to analyze the details of the SM contribution in order to
understand what should be the features of CSM.\\

References about previous works on this process are given in
\cite{ggZH}, in \cite{ggVH} and analytic
expressions can be found in \cite{Kniehl}.\\

The SM amplitudes arise from the quark loop diagrams depicted in Fig.1;
triangle diagrams (a),(b) and box diagrams (c); gluon-gluon symmetrization
is always applied.\\

The leading contributions come from top quark loops in (a) and (c)
whereas the (b) contribution is much smaller. This smallness arises from
the mass suppressed $ZZH$ coupling appearing in (b) as compared to the $G^0ZH$
one in (a) which has in addition the ($m_t/m_W$) enhancement due to the
$G^0tt$ coupling.\\

We will discuss the helicity amplitudes  $F_{\lambda,\lambda',\tau}$
for $g,g,Z$ helicities $\lambda,\lambda' =\pm{1\over2}$ and $\tau=0,\pm1$.\\

The triangle diagrams (a) and (b) only contribute to the $F_{\pm\pm0}$
amplitudes whereas the boxes (c) contribute to all amplitudes.\\

The real and imaginary parts of the various amplitudes
satisfying CP conservation and Bose statistics (see \cite{ggZH})

\bq
F_{---}=F_{+++} ~~~~F_{--0}=-F_{++0}~~~~ F_{--+}=F_{++-}
\eq
\bq
F_{-+-}=F_{+-+}~~~~  F_{-+0}=-F_{+-0}~~~~F_{-++}=F_{+--}
\eq

\bq
F_{\lambda,\lambda',\tau}(\theta)=(-1)^{\tau}F_{\lambda',\lambda,\tau}(\pi-\theta)
\eq
\noindent
are shown in Fig.2,3 in the low energy and in the high energy domains.\\
One can see that the helicity concerving (HC) amplitudes $F_{\pm\mp0}$
finally dominate the helicity violating (HV) ones at high energy
in agreement with the HC rule \cite{hc}.\\

A very important feature is at the origin of the behaviour of the $F_{\pm\pm0}$
amplitudes. It results from very large and opposite contributions of diagrams (a)
and (c), whose peculiar cancellation is illustrated
in Fig.4.\\

One can now check gauge invariance and the Goldstone equivalence of this
property. The same result is indeed
obtained by using the unitary gauge for the $Z$ echange between the quark triangle
and the final $ZH$. The $gg\to Z_LH$ amplitudes also agree, up to $m^2_Z/s$ corrections,
with the $gg\to G^0H$ ones. These amplitudes can be obtained from similar triangle
and box diagrams (replacing $Z$ by $G^0$); however the equivalence does not
appear diagram by diagram,
but only for the total as required by gauge invariance.
In particular for $gg\to G^0H$ the triangle diagrams are now immediately suppressed
whereas the dominant contribution only arises from the box terms.\\

The above properties will play an important role when modifications of the Higgs and top
quark couplings originating from compositeness will be considered in the next
Section.\\

Before, let us show, in Fig.5-7, how the above amplitudes contribute to the unpolarized
and longitudinally polarized Z cross sections and to their energy and angular distributions.\\

Below 1 TeV the dominant amplitudes are the $F_{\pm\pm0}$ (HV) ones,
but after the occurence of the  typical cancellation, above 1 TeV, these amplitudes
are progressively suppressed and the dominant ones become the $F_{\pm\mp0}$ (HC).
The energy distribution of the total cross section is controlled by this property
and indeed is determined at low energy by the HV and at high energy by the HC
amplitudes.\\
The $Z_L$ production rate $P(Z_L)=\sigma(Z_L)/\sigma$
follows this change and becomes close to 1 at high
energy.\\

The shape of the angular distribution results also from the above cancellation.
The constant distribution arising from the triangle (a) is cancelled by a similar
one coming from the boxes (c) and the resulting typical (symmetrical due to Bose
statistics) shape due to the leading HC amplitudes is shown in Fig.7
at 4 TeV (but is rather similar for any other energy).\\
We can anticipate the occurence of large constant distributions when
the above cancellation would be perturbed.\\

\section{COMPOSITE STANDARD MODEL CONSTRAINTS}

The concept of CSM consists in assuming that compositeness (possibly
substructures) preserves the SM properties at low energy
but progressively reveals its presence at high energy, for example
through the effect of form factors but without anomalous couplings
modifying their structure.\\

In the preceding Section, we have seen that the main high energy features
of the $gg\to ZH$ process are the dominance of $Z_LH$ production with
leading amplitudes being the HC ones $F_{\pm\mp0}$, but that this happens
because the suppressed (HV) $F_{\pm\pm0}$ amplitudes result from
the cancellation of very large contributions arising from diagrams (a) and (c).\\

If one introduces form factors in ($G^0Z_LH$) and ($ZZ_LH$)
couplings, appearing only in diagrams (a),(b), this destroys,
as shown in \cite{ggZH}, the cancellation with the boxes ($c$) which do not
involve these couplings and leads to a very
large increase of the $F_{\pm\pm0}$ amplitudes and of the cross sections.\\

Thus, an a priori minor modification immediately violates the CSM expectation. We
would like to see if it could be restored in some way.
We show below that one possibility could be to add top compositeness
and to introduce adequate form factors.\\

There are several options corresponding to top compositeness
affecting only $t_R$ or both $t_L$ and $t_R$, \cite{partialcomp}.
Also the top quark may have common constituents with the Goldstone
and Higgs bosons.
These various options control the details of the corresponding form factors.
We will treat them in a purely phenomenological way by
affecting  form factors to the top couplings appearing
in the triangle and box diagrams.\\

This may be the purely effective description of a complex situation involving
additional heavy quarks or quark constituents with
non perturbative effects due to binding.
We will assume that this can be globally described by introducing scale (s)
dependent modifications of the couplings represented by adequate form factors.\\

Incidently we note that gluon-$tt$ form factors may also appear
but as they would occur exactly in the same way
in the triangles and in the boxes, they would not affect the structures
of the amplitudes (in particular the special cancellation mentioned above)
so we do not discuss them more; they would only modify the total result
by a pure normalization factor.\\

In addition to Higgs form factors for the $G^0Z_LH$ and $ZZ_LH$ vertices
(introduced in  \cite{ggZH})
we now affect the following form factors  to the top couplings:\\

$Htt$:  $g_{Htt}F_{Htt}(s)$ .\\

$Ztt$:  $g^Z_{tL}P_LF_{tL}(s)+g^Z_{tR}P_RF_{tR}(s)$ (left and right form factors).\\

$G^0tt$: $g_{G^0tt}\gamma^5F_{Gtt}(s)$.\\

We obtain the following results for the 3 types of diagrams:\\
 
--- the triangle (a) with factor $g_{G^0Z_LH}F_{G^0Z_LH}(s)g_{G^0tt}F_{Gtt}(s)$ ;\\

--- the triangle (b) with factor
$g_{ZZH}F_{ZZ_LH}(s)(g^Z_{tL}P_LF_{tL}(s)+g^Z_{tR}P_RF_{tR}(s)$) ;\\

--- the boxes (c) with factor
$g_{Htt}F_{Htt}(s)(g^Z_{tL}P_LF_{tL}(s)+g^Z_{tR}P_RF_{tR}(s))$.\\
However, as in the SM case, the addition of the direct, crossed and twisted boxes leaves
only the R-L combination. This means a result proportional to
$g_{Htt}F_{Htt}(s)(g^Z_{tR}F_{tR}(s)-g^Z_{tL}F_{tL}(s))$.\\

So in general, with five arbitrary form factors
$F_{G^0Z_LH}(s)=F_{ZZ_LH}(s)$, $F_{Gtt}(s)$, $F_{Htt}(s)$, $F_{tR}(s)$, $F_{tL}(s)$,
the cancellation of (a) and (c) does not occur.\\

This cancellation can be recovered provided that
\bq
g_{G^0Z_LH}F_{G^0Z_LH}(s)g_{G^0tt}F_{Gtt}(s)=
g_{Htt}F_{Htt}(s)(g^Z_{tR}F_{tR}(s)-g^Z_{tL}F_{tL}(s))~~\label{CSMcons}
\eq
Note that this equality is indeed satisfied when the form factors are equal to 1,
as we have
\bq
g_{G^0Z_LH}g_{G^0tt}=g_{Htt}(g^Z_{tR}-g^Z_{tL})={m_t\over4s^2_Wc_Wm_W}
\eq
So we can rewrite the general constraint among the five form factors
as follows:\\

\underline{CSM constraint}:
\bq
F_{G^0Z_LH}(s)F_{Gtt}(s)(g^Z_{tR}-g^Z_{tL})=
F_{Htt}(s)(g^Z_{tR}F_{tR}(s)-g^Z_{tL}F_{tL}(s))~~\label{CSMconsFF}
\eq

This connection between compositeness of the top and of the Higgs sectors
is rather peculiar and we will consider some specific applications.\\

--- (a) For simplicity we may assume that the Higgs boson structure gives a unique
form factor $ F_H(s)$:
\bq
F_{G^0Z_LH}(s)=F_{Htt}(s)\equiv F_H(s)
\eq
in this case the CSM constraint reduces to

\bq
F_{Gtt}(s)(g^Z_{tR}-g^Z_{tL})=
(g^Z_{tR}F_{tR}(s)-g^Z_{tL}F_{tL}(s))~~\label{CSMtLR}
\eq

It appears very natural for a CSM
description which should preserve
the Goldstone equivalence and therefore impose the same modification for
$G^0$ and for $Z$ couplings.\\
In this case we have 3 independent form factors
$F_H(s)$, $F_{tR}(s)$,$F_{tL}(s)$ and
we can use this CSM constraint as it stands
directly for the case of both $t_L$ and $t_R$ compositeness,\\

--- (b) In the case where only $t_R$ is affected by compositeness one gets
\bq
F_{tL}(s)\equiv 1~~~~F_{Gtt}(s)(g^Z_{tR}-g^Z_{tL})=(g^Z_{tR}F_{tR}(s)-g^Z_{tL})~~\label{CSMtR}
\eq
and 2 independent form factors, $F_H(s)$ and $F_{tR}(s)$.\\

As examples for the illustrations we will consider simpler "test" cases by using
only one form factor shape and imposing respectively\\

--- (a') in the $t_{L,R}$ compositeness case $F_H(s)=F_{tR}(s)=F_{tL}(s)$,
denoted CSMtLR,\\
 
--- (b') in the pure $t_R$ compositeness case $F_H(s)=F_{tR}(s)$ and $F_{tL}(s)=1$
denoted CSMtR.\\

All these cases satisfy the CSM constraint that we
retain as a basic property for the $gg\to ZH$ process.
In the next Section and Figures 8-13, we illustrate the consequences of such choices
for the amplitudes and cross sections
and we compare them with other compositeness cases which violate this constraint.\\

\section{ILLUSTRATIONS}

We compare the SM predictions to two CSM conserving cases CSMtLR, CSMtR and
two CSM violating cases called CSMvt (with top and Higgs form factors)
and CSMvH (with only Higgs form factors).\\

In the following illustrations we use for simplicity the expression for the form factors:
\bq
F(s)={(m_Z+m_H)^2+M^2\over s+M^2}
\eq
\noindent
with the new physics scale taken for example as $M=5$ TeV. Its aim is just to show what type of modifications
such form factors could generate, but no to make a definite prediction.\\

We illustrate 

---- case CSM (a') with
$F_{Htt}(s)=F_{tR}(s)=F_{tL}(s)\equiv F(s)$,
and $F_{Gtt}(s)$ determined by (\ref{CSMtLR}),\\
 
---- and CSM case (b') with $F_{Htt}(s)=F_{tR}(s)\equiv F(s)$, $F_{tL}(s)=1$,
and $F_{Gtt}(s)$ determined by (\ref{CSMtR}),\\

For comparison we will illustrate the 2 CSM violating cases:\\

---- (c) denoted (CSMvt) where we use $F_{tR}(s)$ with $M=10$ TeV, $F_{tL}(s)$ 
with $M=15$ TeV
and 
$F_{Htt}(s)=F_{Gtt}(s)\equiv F(s)$ with $M=5$ TeV, \\

---- (d) denoted (CSMvH) without top form factor and only one  $F_H(s)=F(s)$ form factor affecting the
$G^0Z_LH$ and $ZZ_LH$ vertices, .\\

Fig.8 shows the consequences for the $F_{++0}$ amplitude. One sees how the CSM violating
cases generate immediately large enhancements of this amplitude because of the
lack of cancellation between (a) and (c), especially in the CSMvH case.\\

Fig.9 shows how the $F_{+-0}$ behaves totally differently and is suppressed at high
energy by the form factors. Note that the CSMvH case does not affect the boxes
and therefore not this amplitude.\\

Fig.10,11 illustrate how these effects lead to the corresponding behaviours
of the cross sections for specific
polarizations $\sigma(\pm\pm0)$ ,$\sigma(\pm\mp0)$, for the unpolarized
cross section $\sigma$ and for the corresponding final $Z_L$ rates.\\

Fig.12,13 show the corresponding angular distributions at 4 TeV of these
various cross sections and of the $Z_L$ rates. One can see how
the CSM violating cases (with the absence of cancellation of the isotropic
triangle (a) (with pure s-channel exchange) contribute to constant angular distributions.\\

Summarizing, spectacular differences appear between CSM conserving and violating
cases with specific behaviours of the various amplitudes and cross sections,
their energy and angular dependences.
The next step  would be to study how these properties could be detected in
hadronic collisions in particular with identification of the $ZH$ final state.\\

\section{CONCLUSIONS}

We have shown that $gg\to ZH$ is a particularly interesting process for testing
the nature of a possible Higgs boson compositeness especially with the CSM concept.
The important role of the top quark in this process
suggests the presence of additional effects arising from top quark
compositeness that could also be constrained by CSM.\\
We have analyzed the behaviour of the amplitudes of the $gg\to ZH$ process
and in particular the peculiar cancellation controlling the HV amplitudes $F_{\pm\pm0}$.
CSM implies a protection of this property.
This requires a peculiar relation between
the two (Higgs and top quark) compositeness sectors and,
explicitly, between the corresponding form factors.
We have examined several possibilities, with $t_L$ and/or $t_R$ compositeness
and we have established the corresponding CSM constraints.
We have illustrated the consequences for the energy and angular dependences of the cross
section and of the percentage of $Z_L$ production.
On another hand we have shown that spectacular modifications with respect to SM
predictions would appear when such CSM relations are violated.\\

We have not considered specific theoretical models and how
the form factors that we introduced would be generated.
They could also be replaced by the exchange of new particles,
for examples heavy resonaces, see ref.\cite{res}.
There are various types of extensions of the SM, see  refs.\cite{Hcomp, comp, Hcomp2, Hcomp3, partialcomp, Hcomp4}.
It could be interesting to see what our CSM constraints would imply for such
models, or if they would suggest new type of models to be constructed, in
particular substructures which could keep the SM features in a simple fashion.\\

Independently, phenomenological studies of possible consequences of
such CSM pictures for other processes may be done.
The case of  photon-photon is under study;
it should be interesting, when using polarized beams, for separating
$F_{\pm\pm0}$ and $F_{\pm\mp0}$ amplitudes which, as we have shown in this
paper, have peculiar properties resulting from the preservation or from the
violation the CSM principle.\\

\newpage

\begin{figure}[p]
\[
\epsfig{file=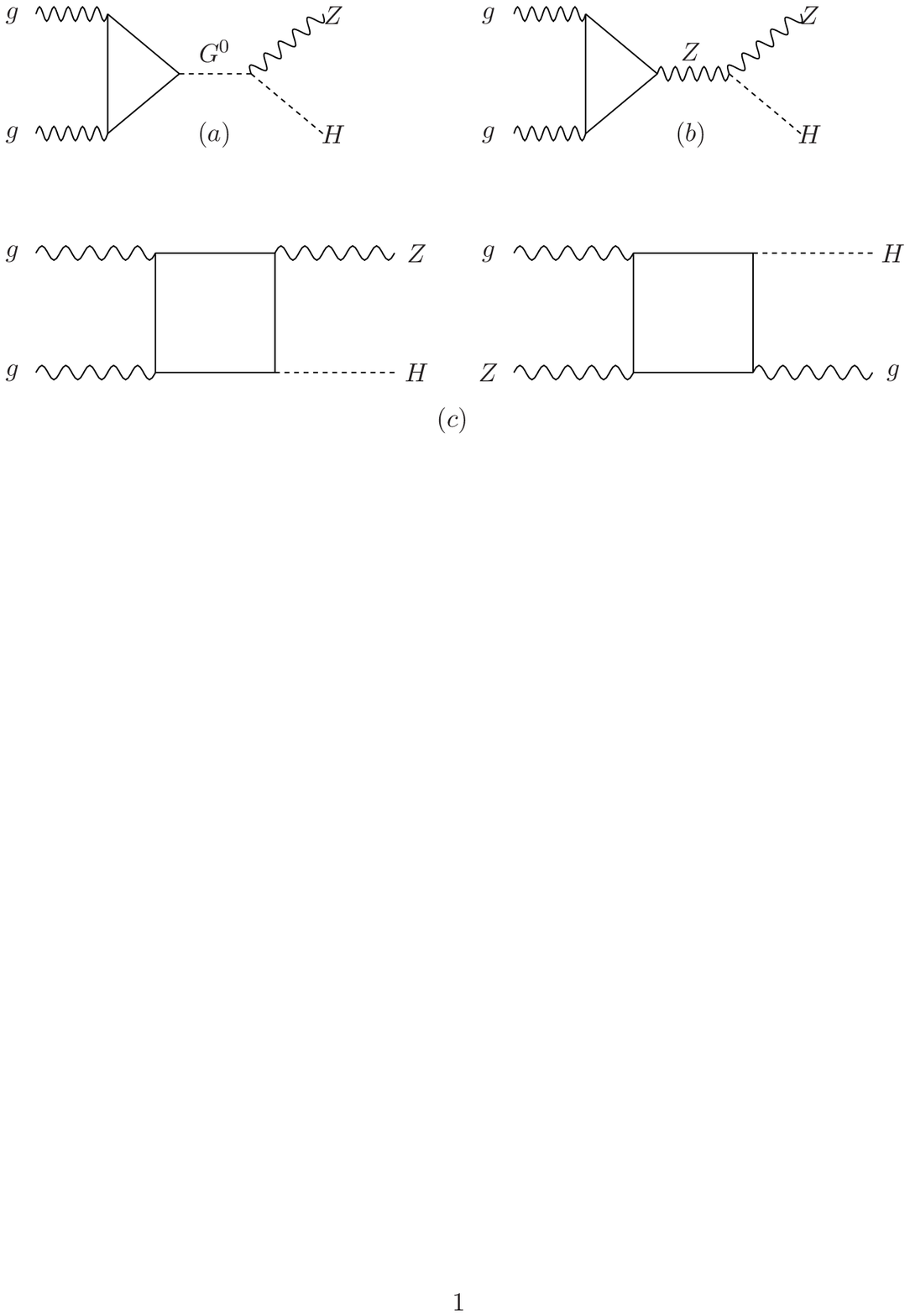, height=24.cm}
\]\\
\vspace{-15cm}
\caption[1] {The triangle and box one loop SM diagrams contributing
to the $gg\to ZH$ process.}
\end{figure}

\clearpage

\begin{figure}[p]
\[
\epsfig{file=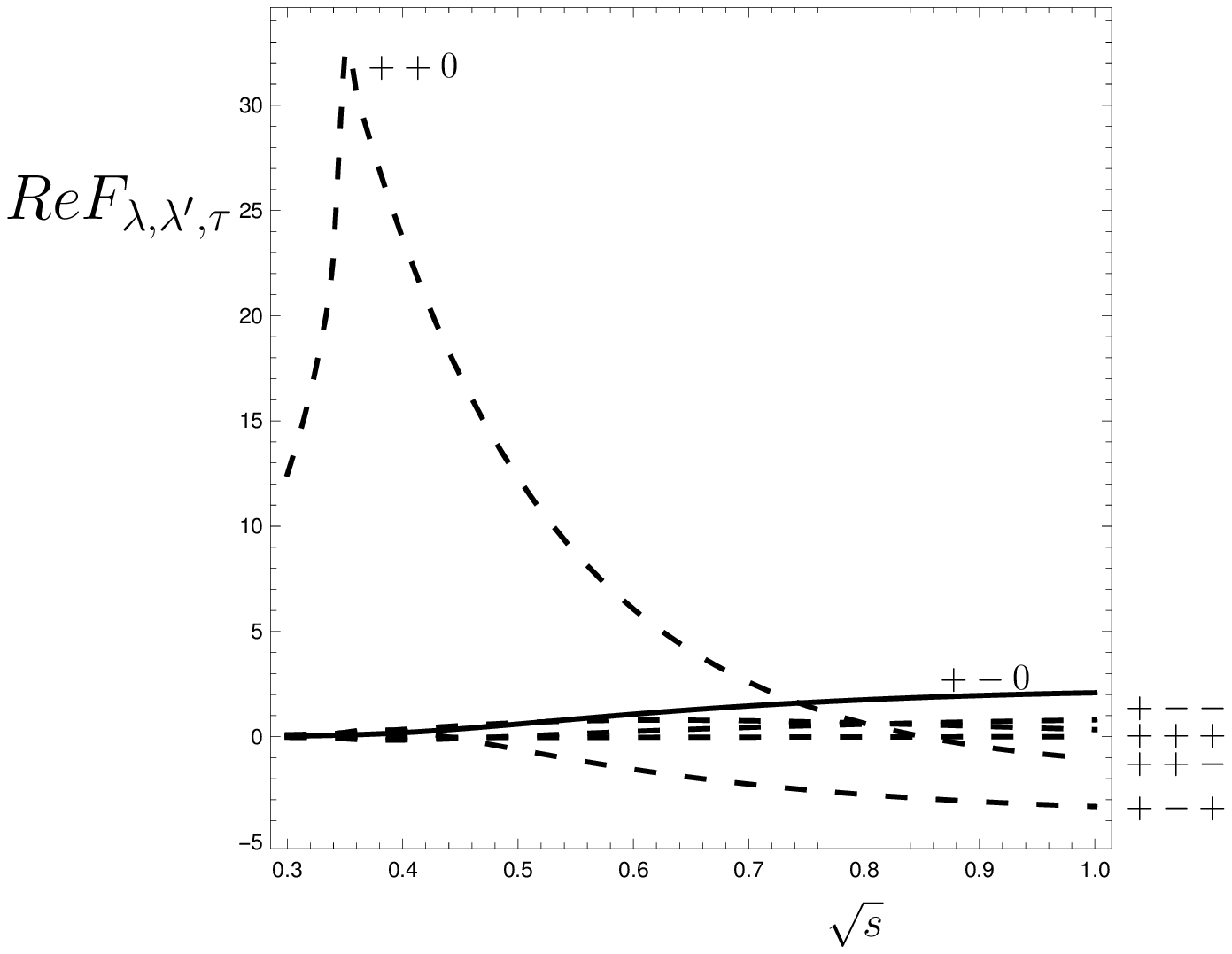, height=8.cm}
\]\\
\vspace{-1cm}
\[
\epsfig{file=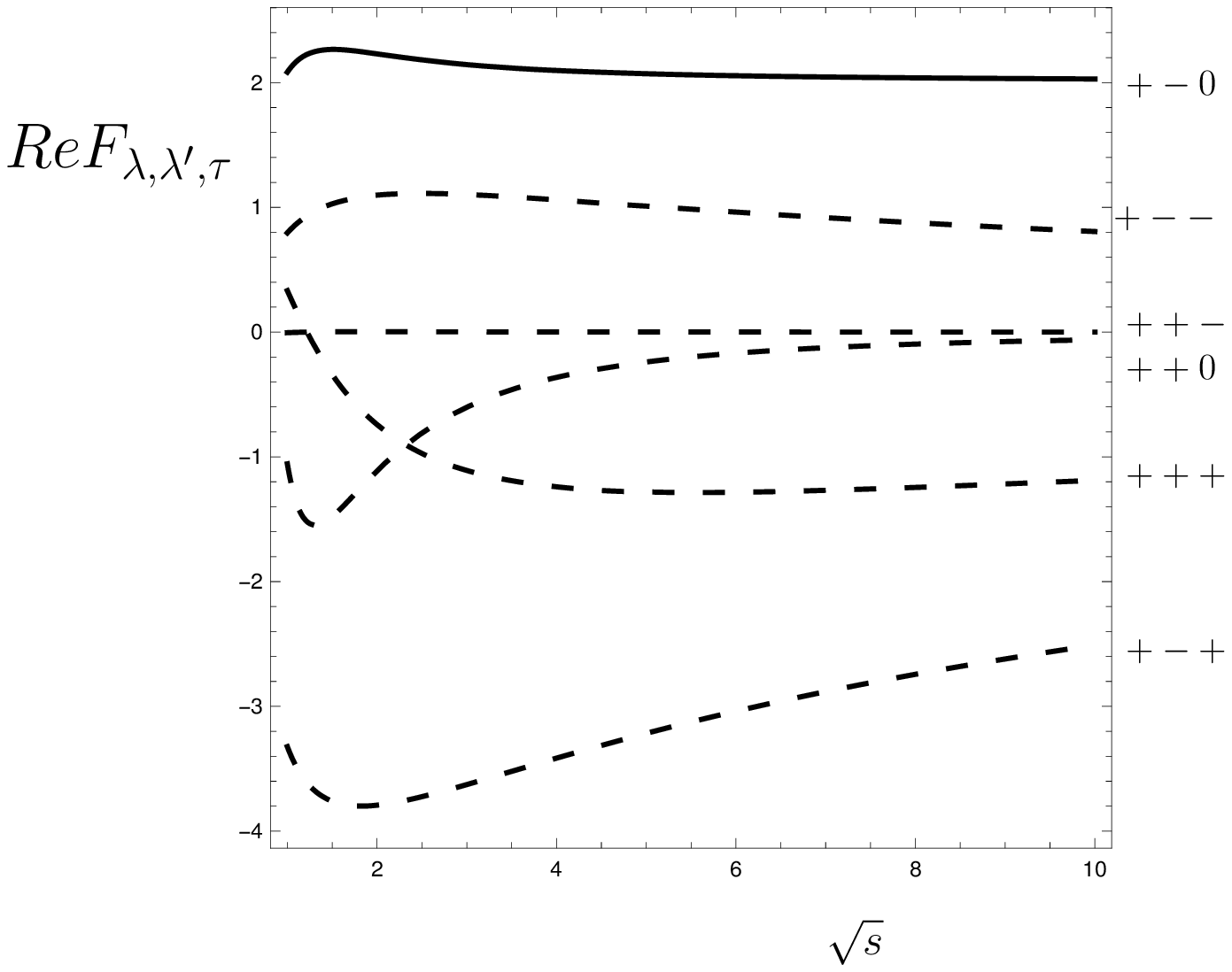, height=8.cm}
\]\\
\vspace{-1cm}
\caption[1] {Energy dependence of the real parts of the 6 independent SM amplitudes.}
\end{figure}

\clearpage

\begin{figure}[p]
\[
\epsfig{file=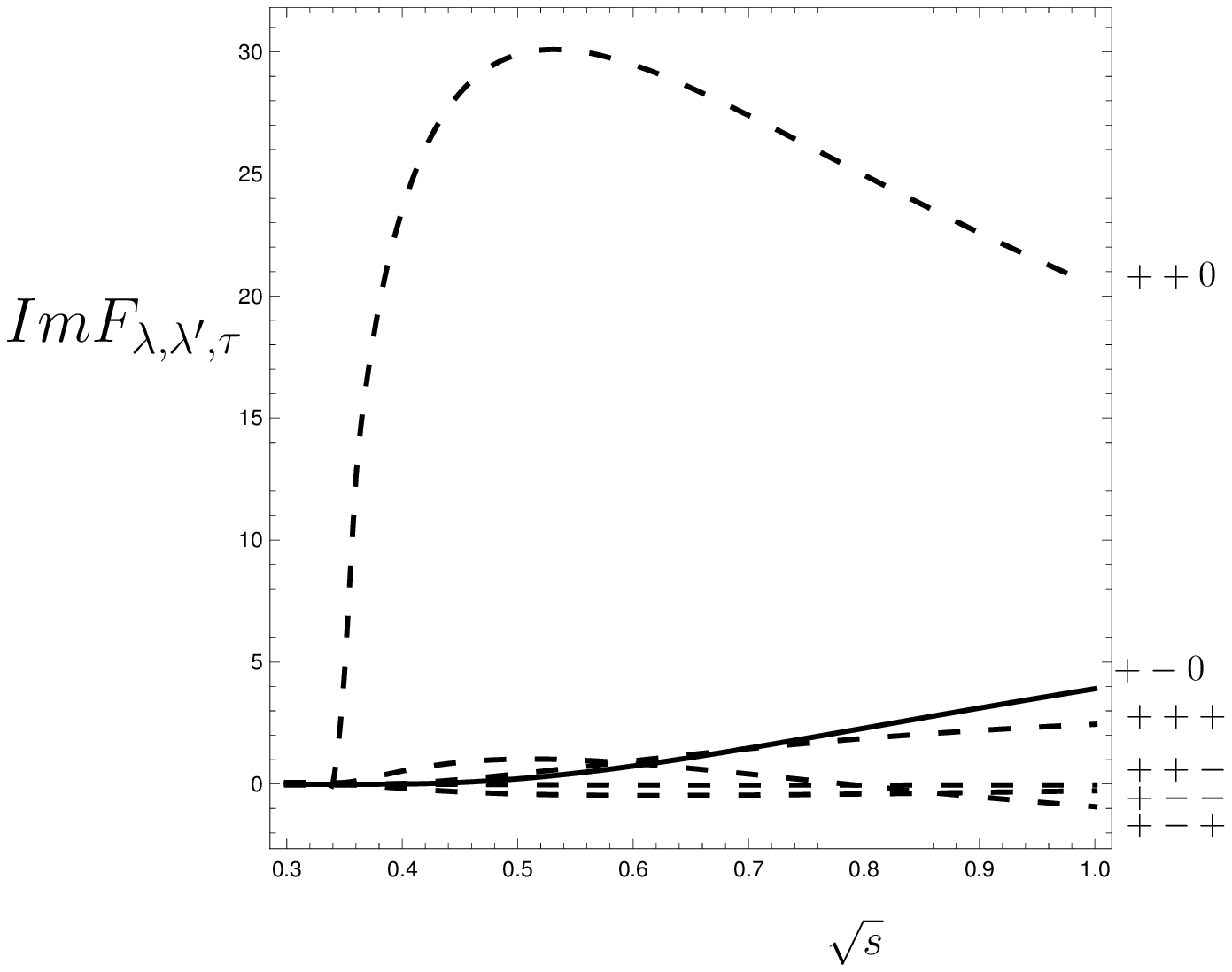, height=8.cm}
\]\\
\vspace{-1cm}
\[
\epsfig{file=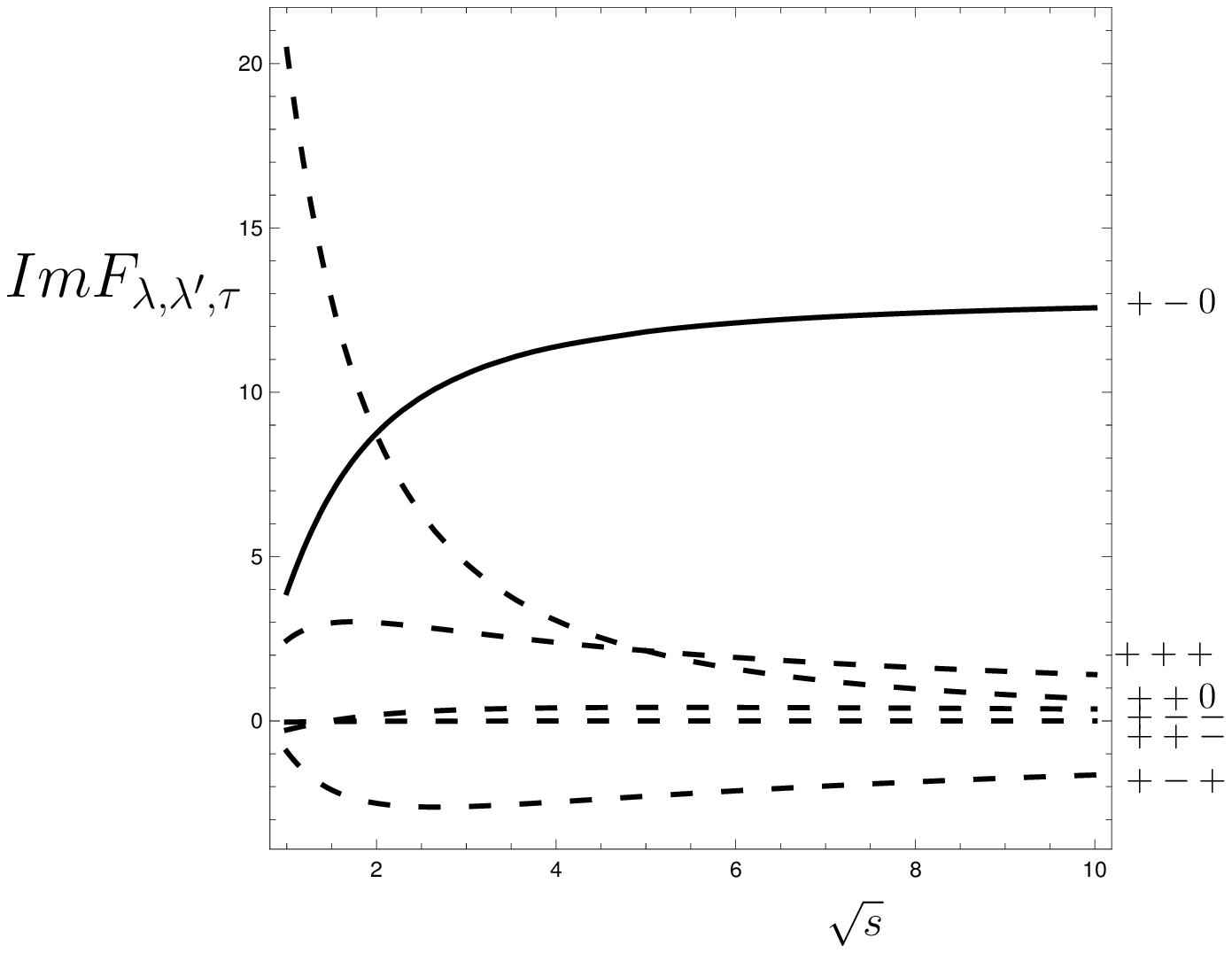, height=8.cm}
\]\\
\vspace{-1cm}
\caption[1] {Energy dependence of the imaginary parts of the 6 independent SM amplitudes.}
\end{figure}

\clearpage

\begin{figure}[p]
\[
\epsfig{file=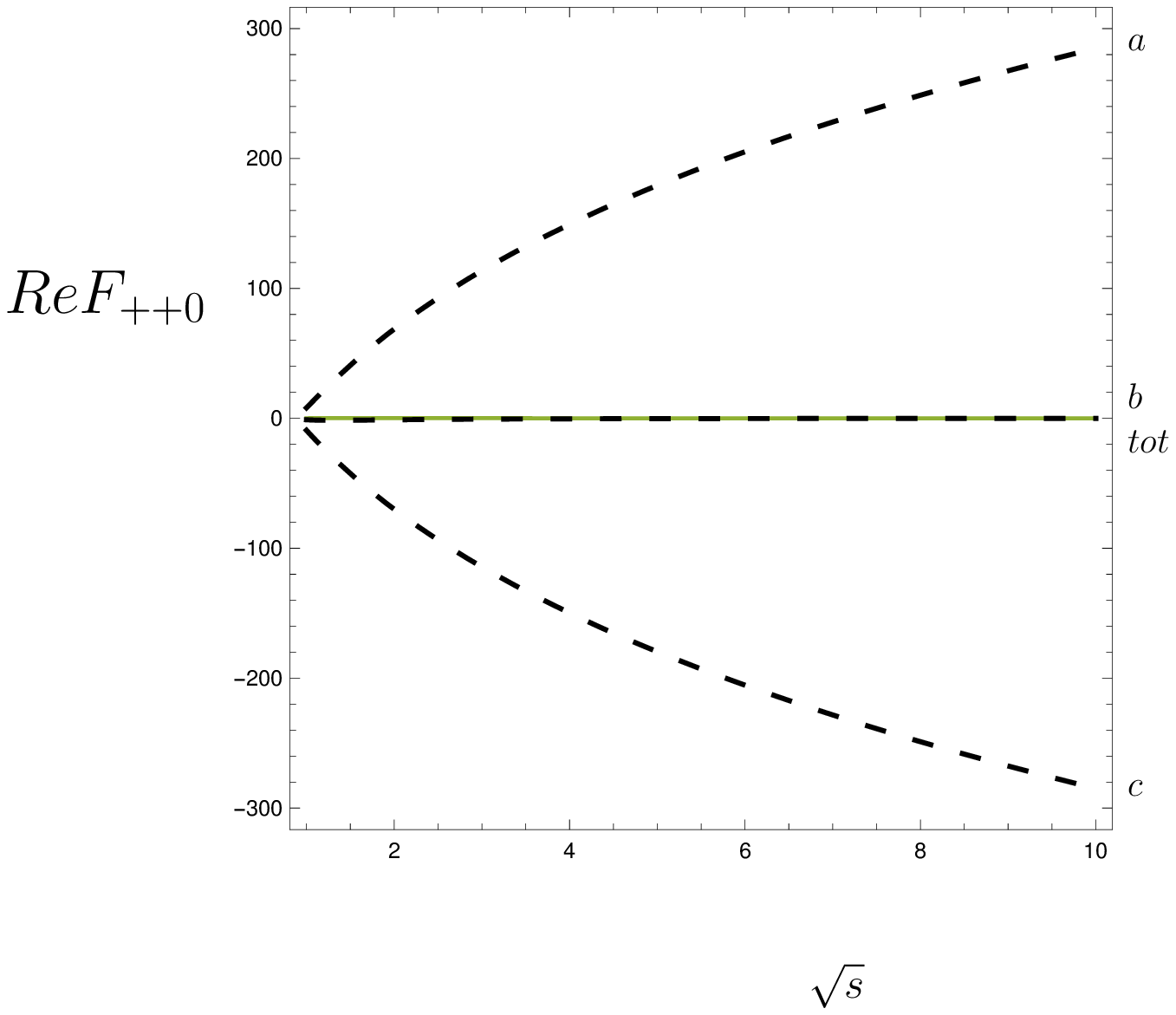, height=8.cm}
\]\\
\vspace{-1cm}
\[
\epsfig{file=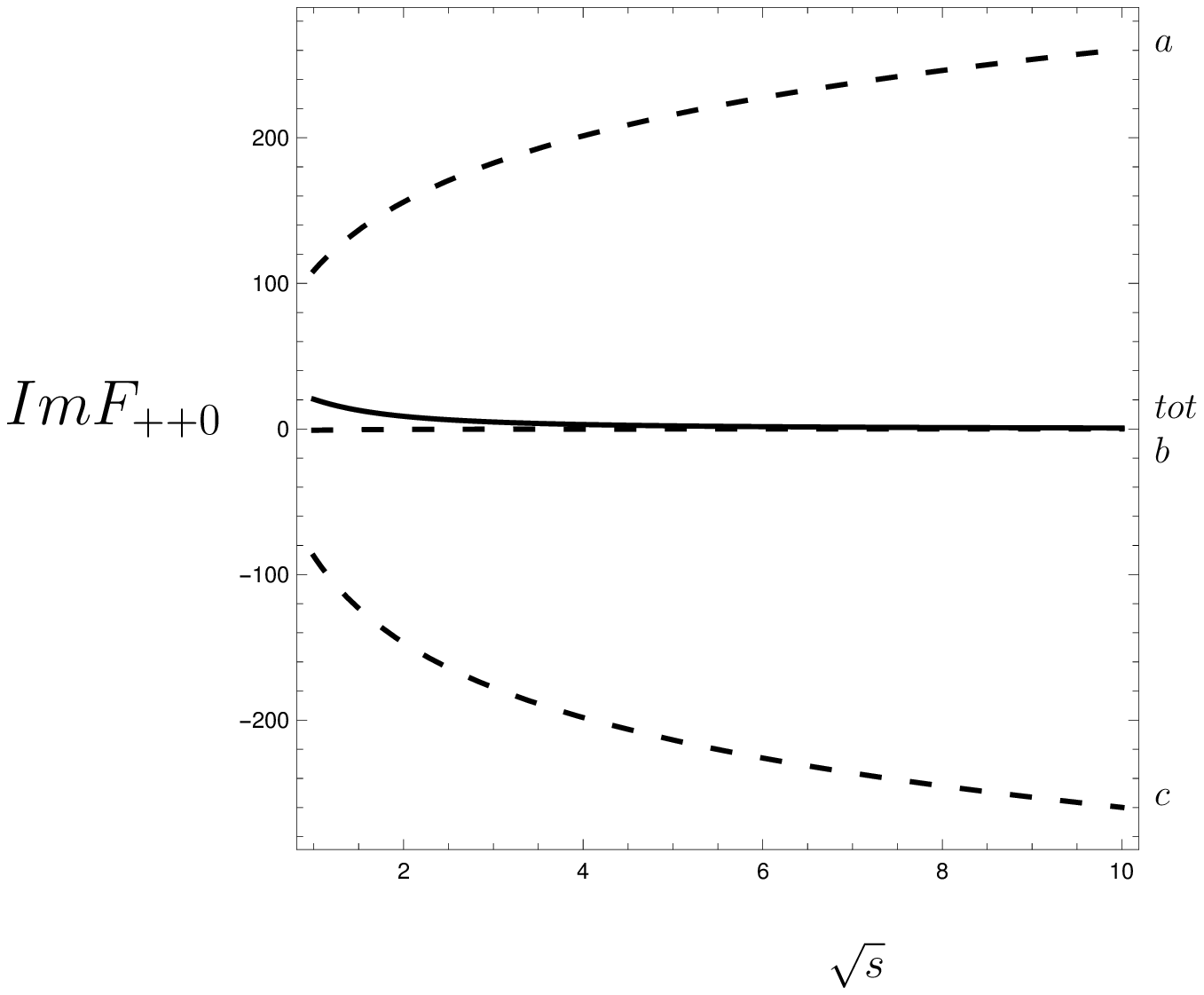, height=8.cm}
\]\\
\vspace{-1cm}
\caption[1] {Energy dependences of the Real and Imaginary parts of the components of the $F_{++0}$ SM amplitude.}
\end{figure}

\clearpage
\begin{figure}[p]
\[
\epsfig{file=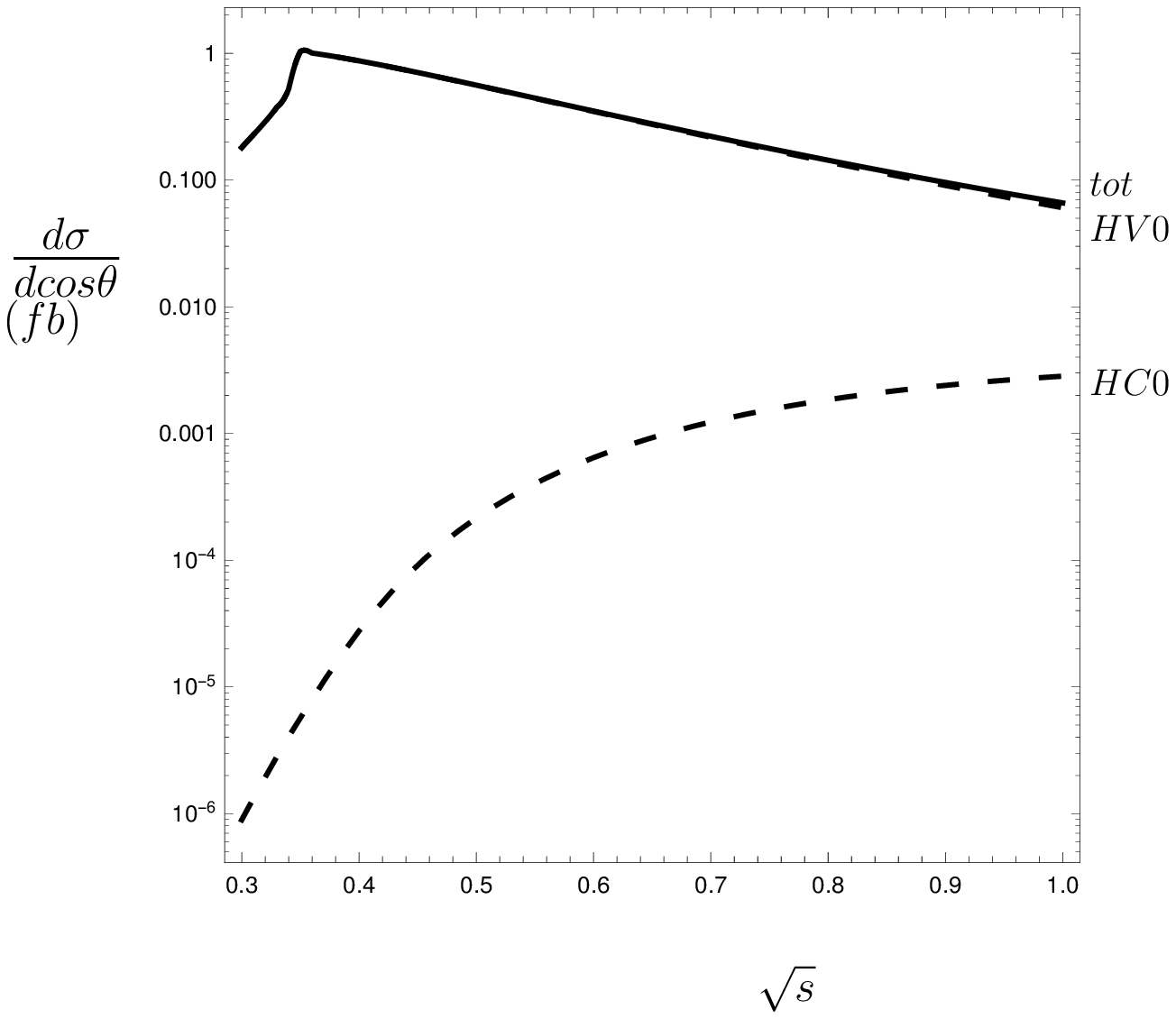, height=8.cm}
\]\\
\vspace{-1cm}
\[
\epsfig{file=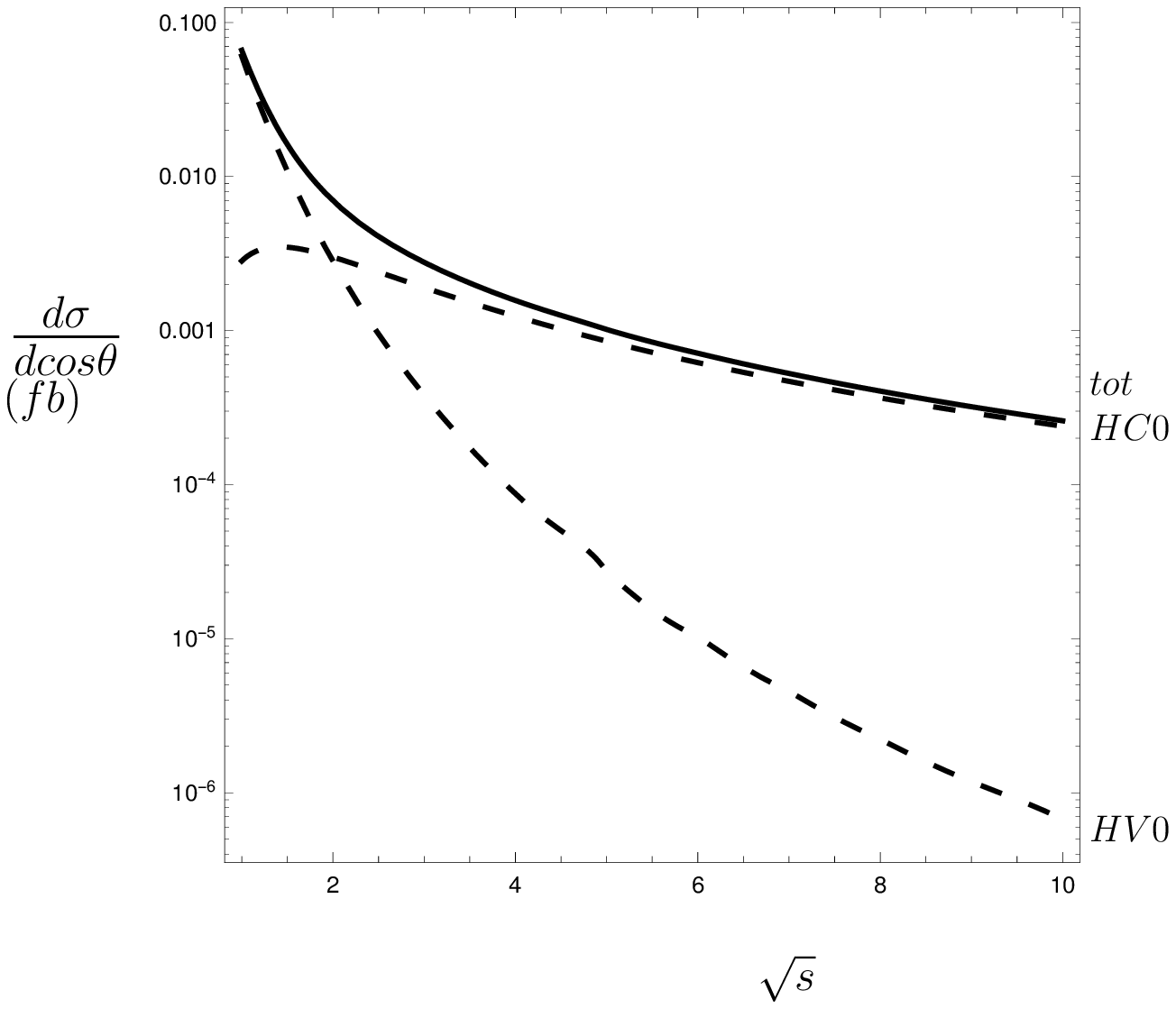, height=8.cm}
\]\\
\vspace{-1cm}
\caption[1] {Energy dependence of the SM cross sections ($\pm\pm0$) denoted HV0,
($\pm\mp0$) denoted HC0, and the unpolarized one denoted tot.}
\end{figure}

\clearpage
\begin{figure}[p]
\[
\epsfig{file=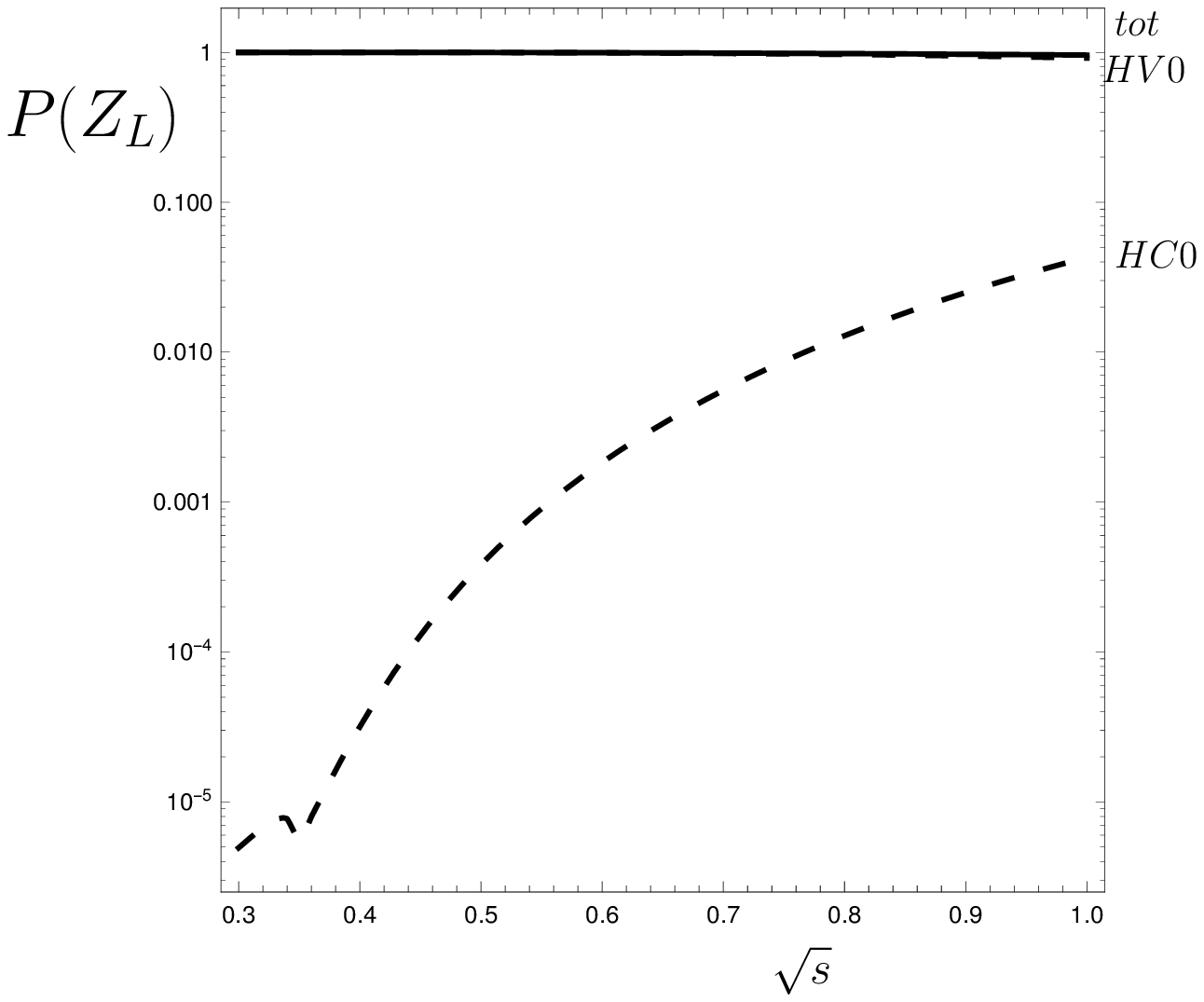, height=8.cm}
\]\\
\vspace{-1cm}
\[
\epsfig{file=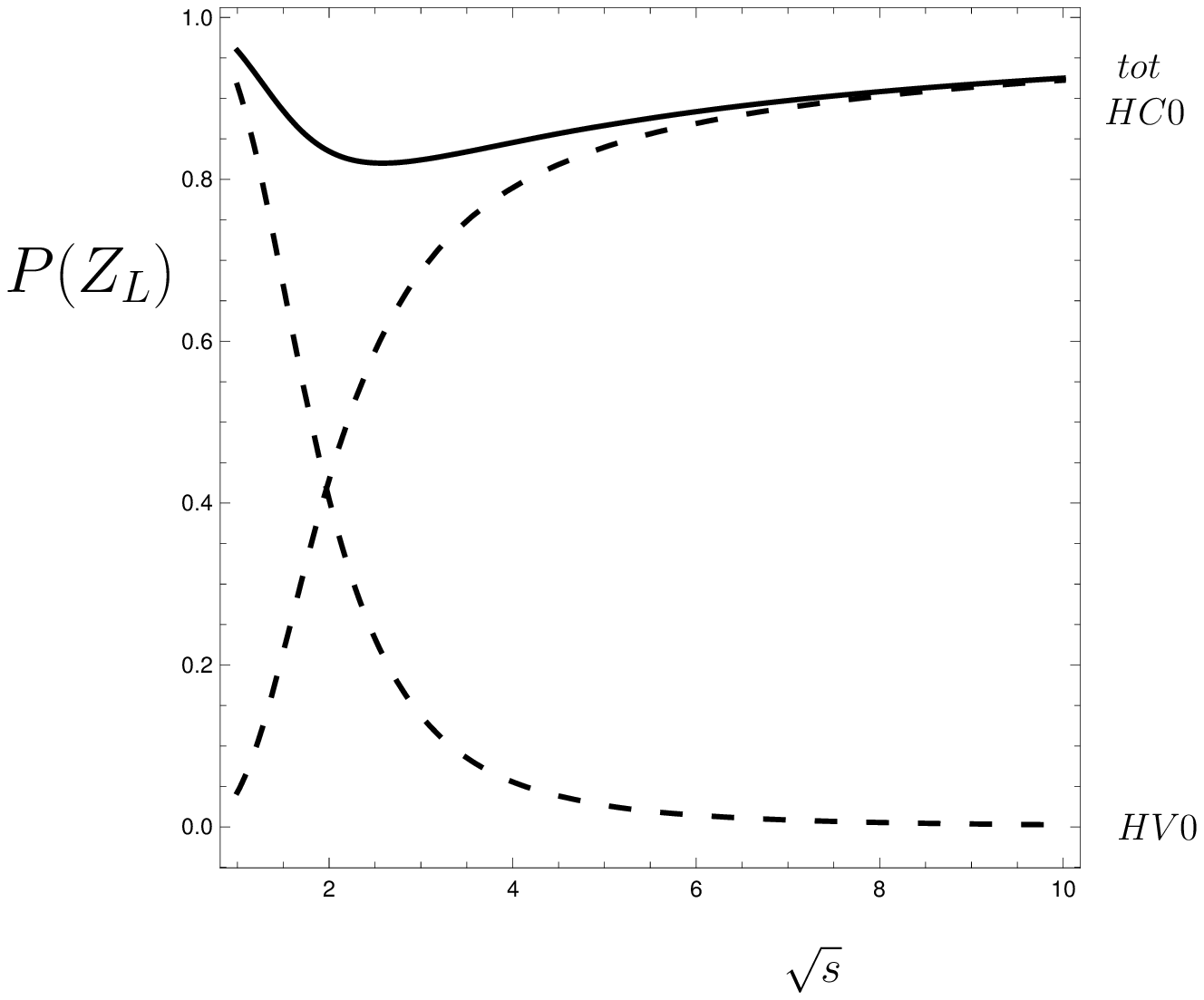, height=8.cm}
\]\\
\vspace{-1cm}
\caption[1] {Energy dependence of the SM longitudinal $Z_L$ production
rate for the same 3 cases as in Fig.5}
\end{figure}

\clearpage
\begin{figure}[p]
\[
\epsfig{file=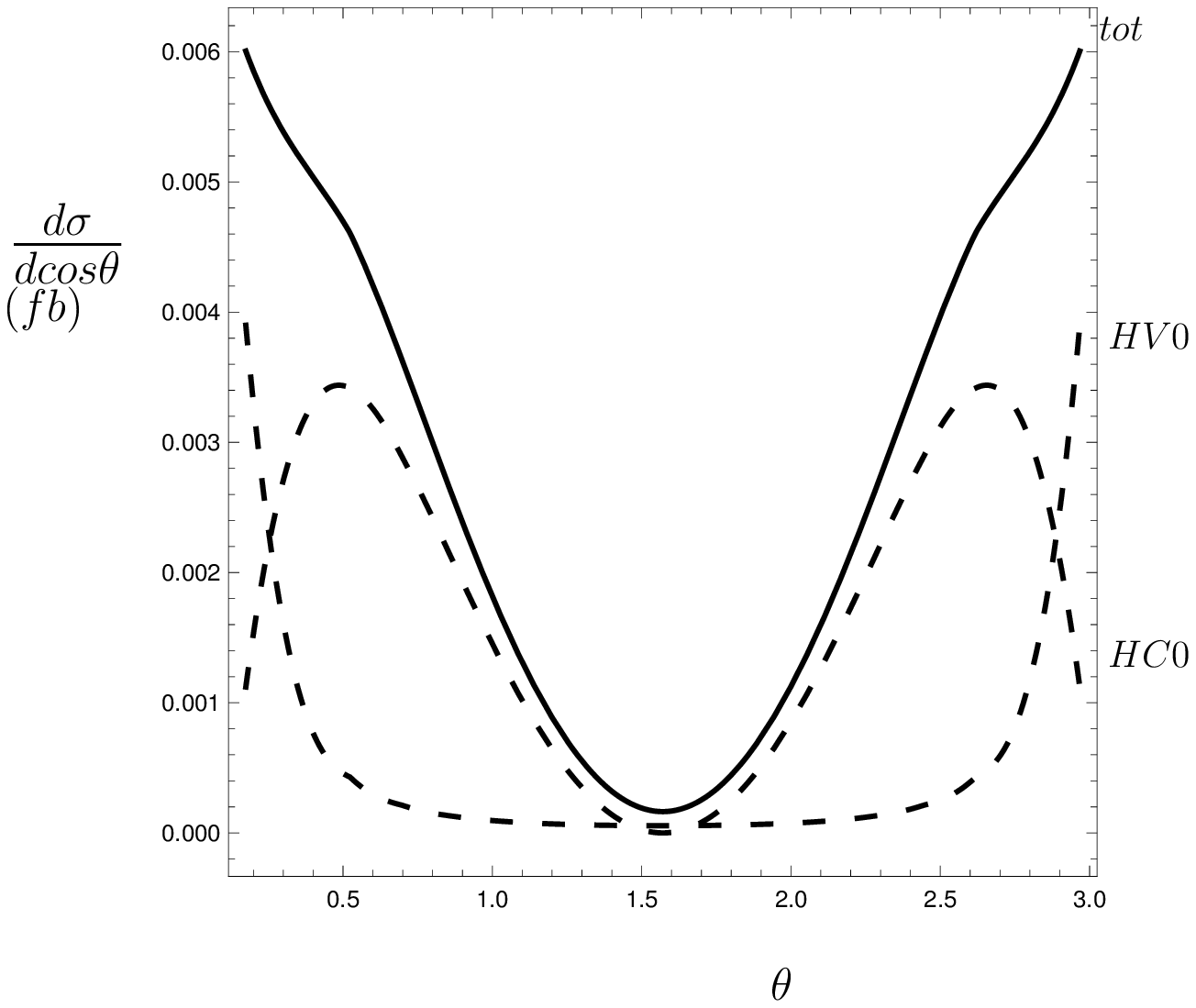, height=8.cm}
\]\\
\vspace{-1cm}
\[
\epsfig{file=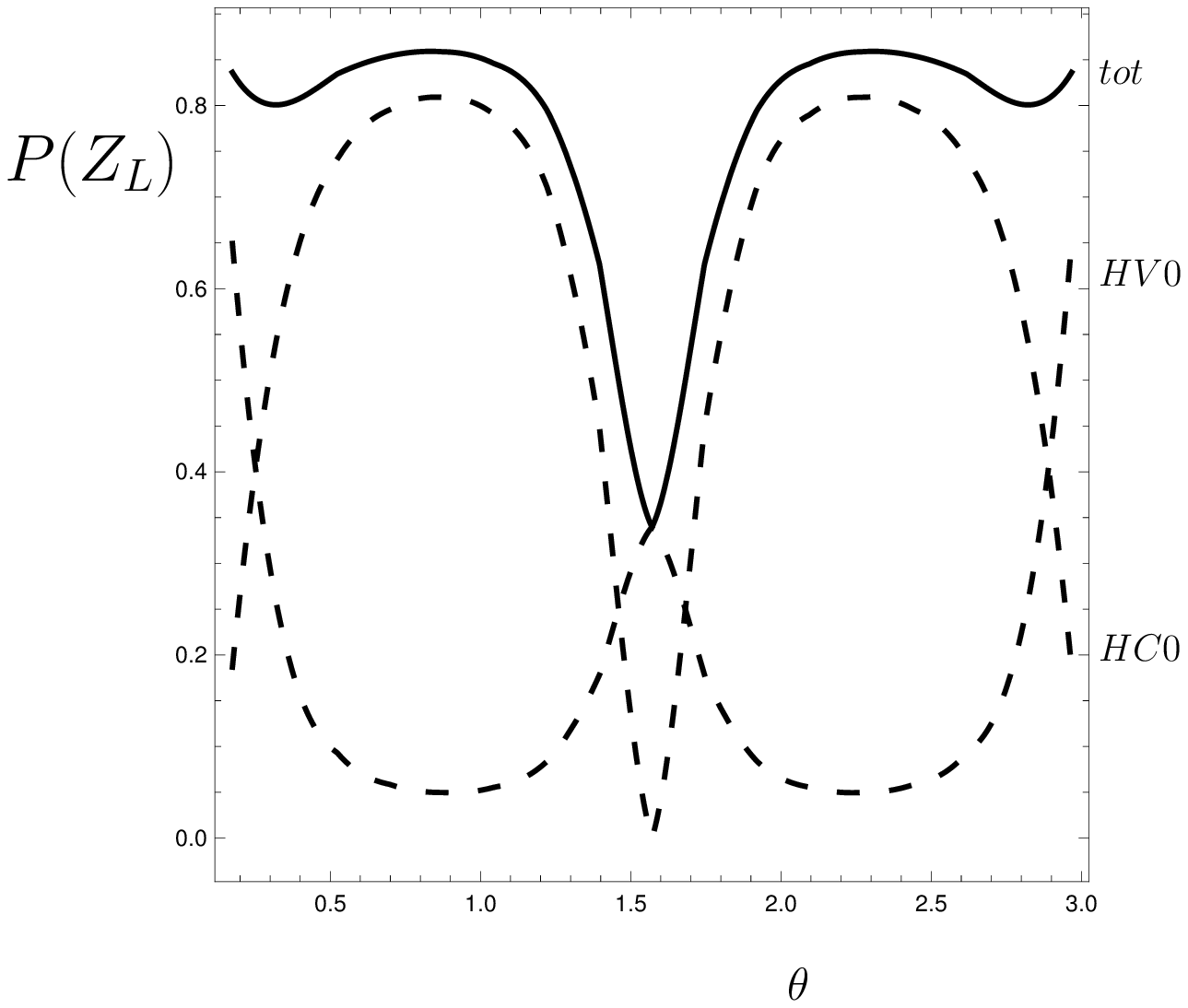, height=8.cm}
\]\\
\vspace{-1cm}
\caption[1] {Angular distributions of SM cross sections  and $Z_L$ rate; same notations as in Fig.5.}
\end{figure}

\clearpage
\begin{figure}[p]
\[
\epsfig{file=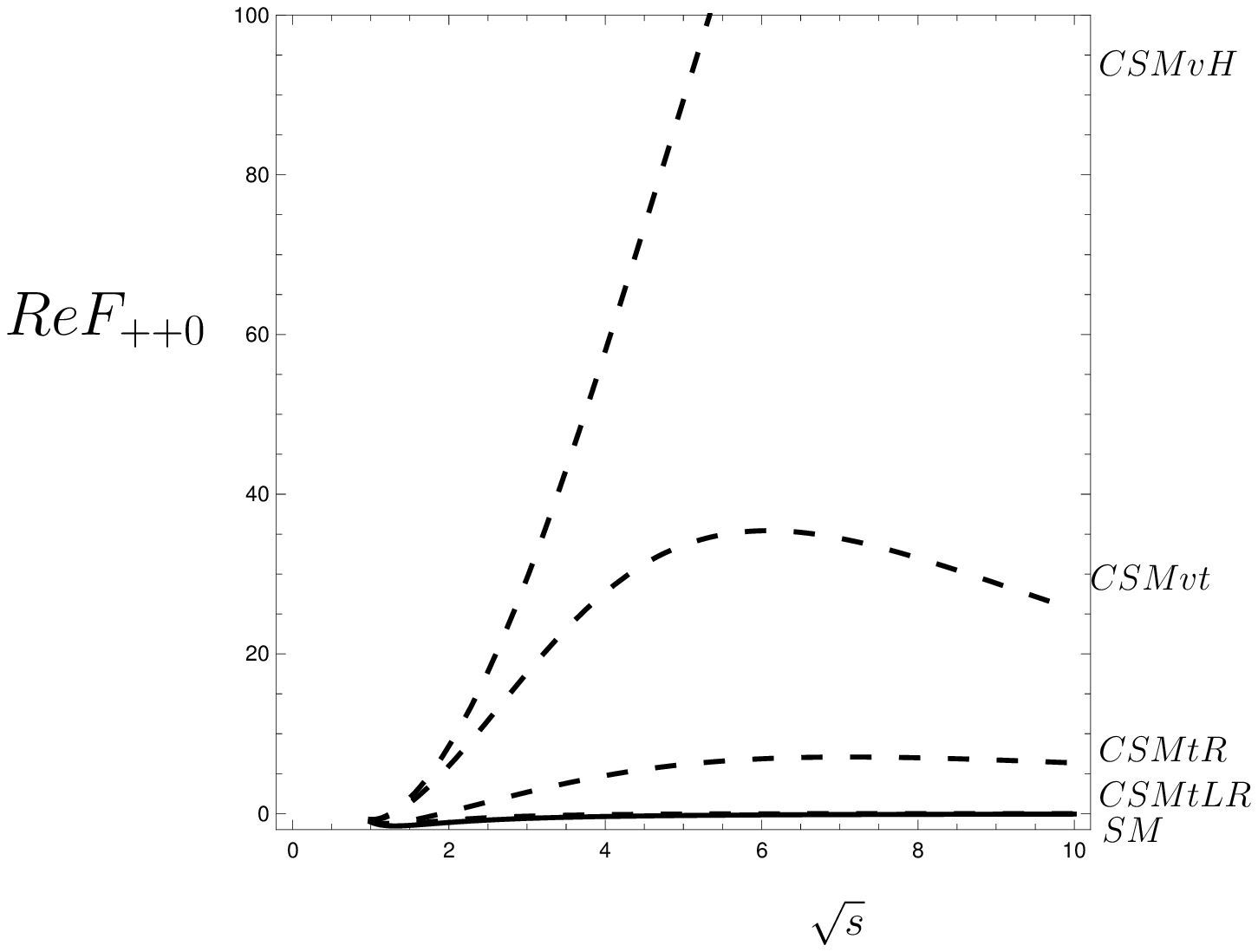, height=8.cm}
\]\\
\vspace{-1cm}
\[
\epsfig{file=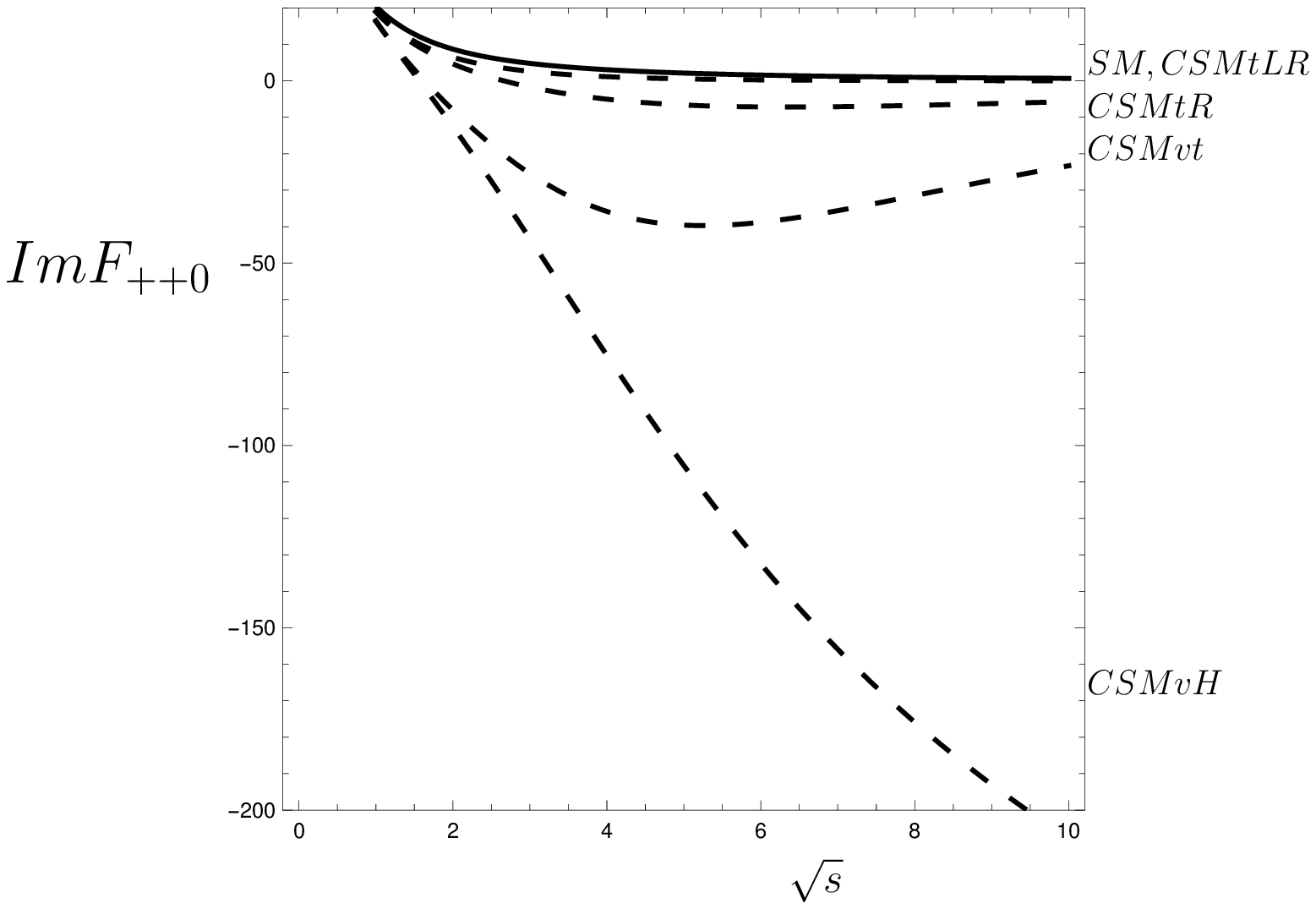, height=8.cm}
\]\\
\vspace{-1cm}
\caption[1] {Real and Im parts of $F_{++0}$ in SM, CSMtLR, CSMtR, CSMvt, CSMvH cases.}
\end{figure}

\clearpage
\begin{figure}[p]
\[
\epsfig{file=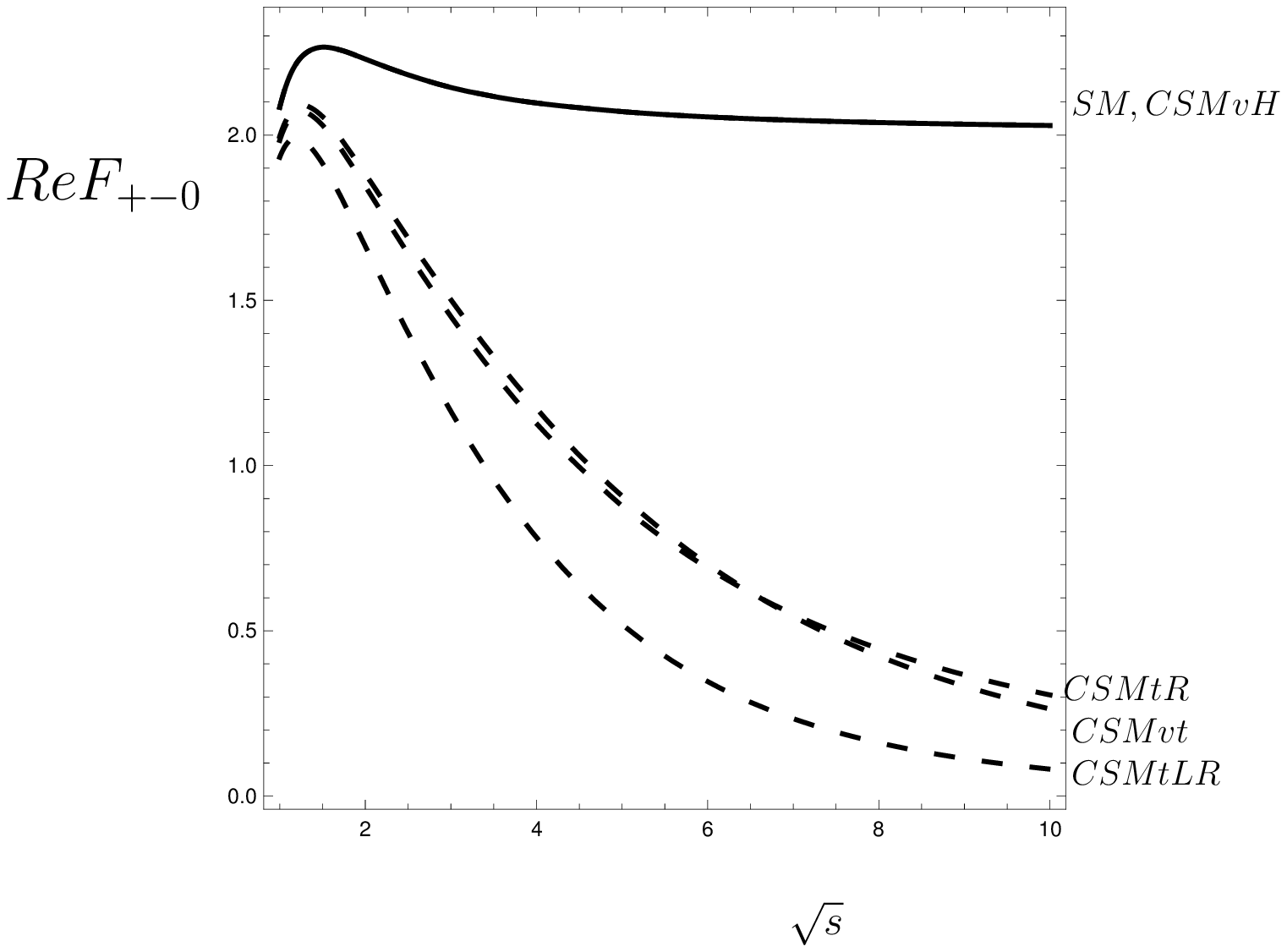, height=8.cm}
\]\\
\vspace{-1cm}
\[
\epsfig{file=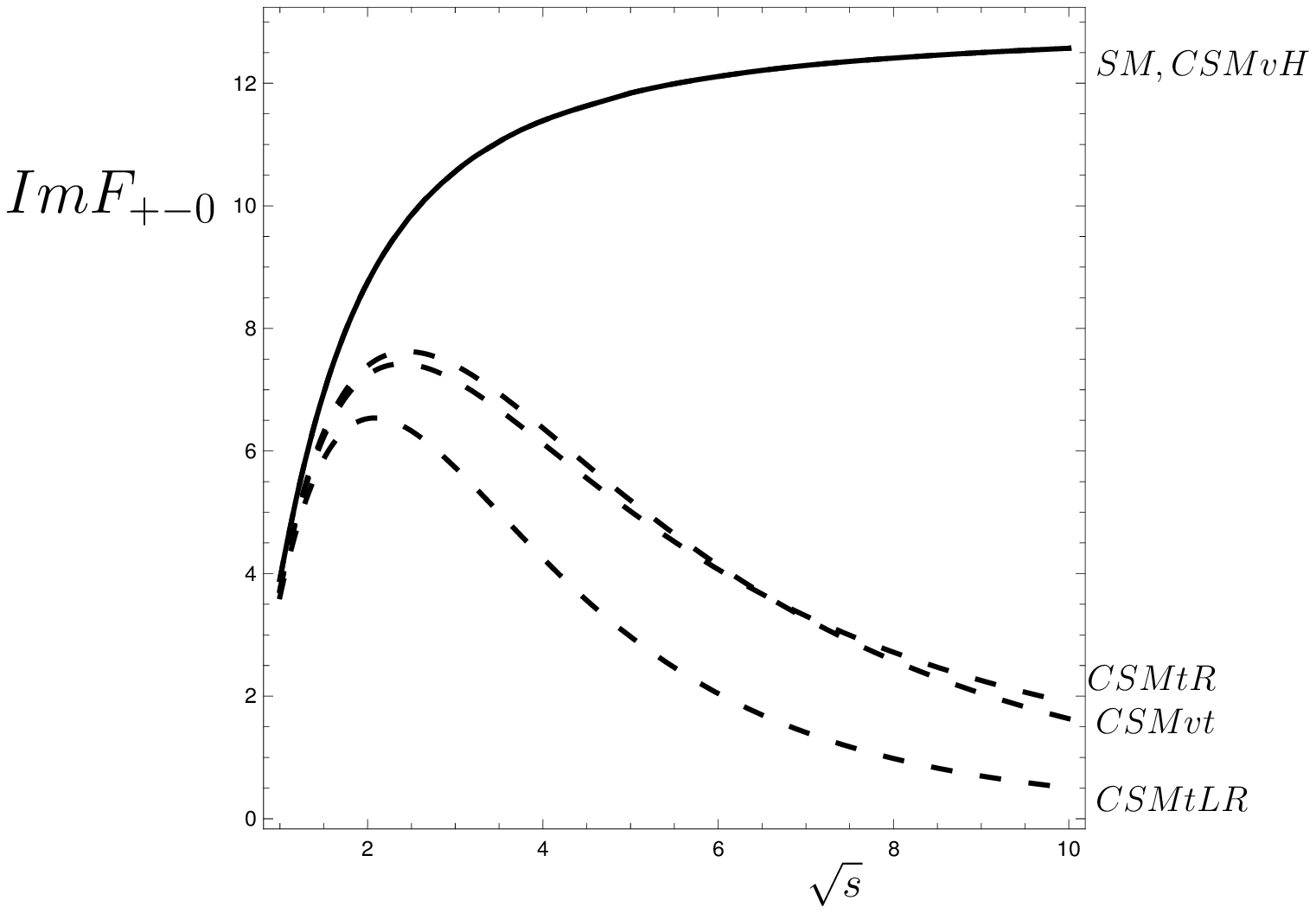, height=8.cm}
\]\\
\vspace{-1cm}
\caption[1] { Real and Im parts of $F_{+-0}$ in SM, CSMtLR, CSMtR, CSMvt, CSMvH cases.}
\end{figure}

\clearpage
\begin{figure}[p]
\[
\epsfig{file=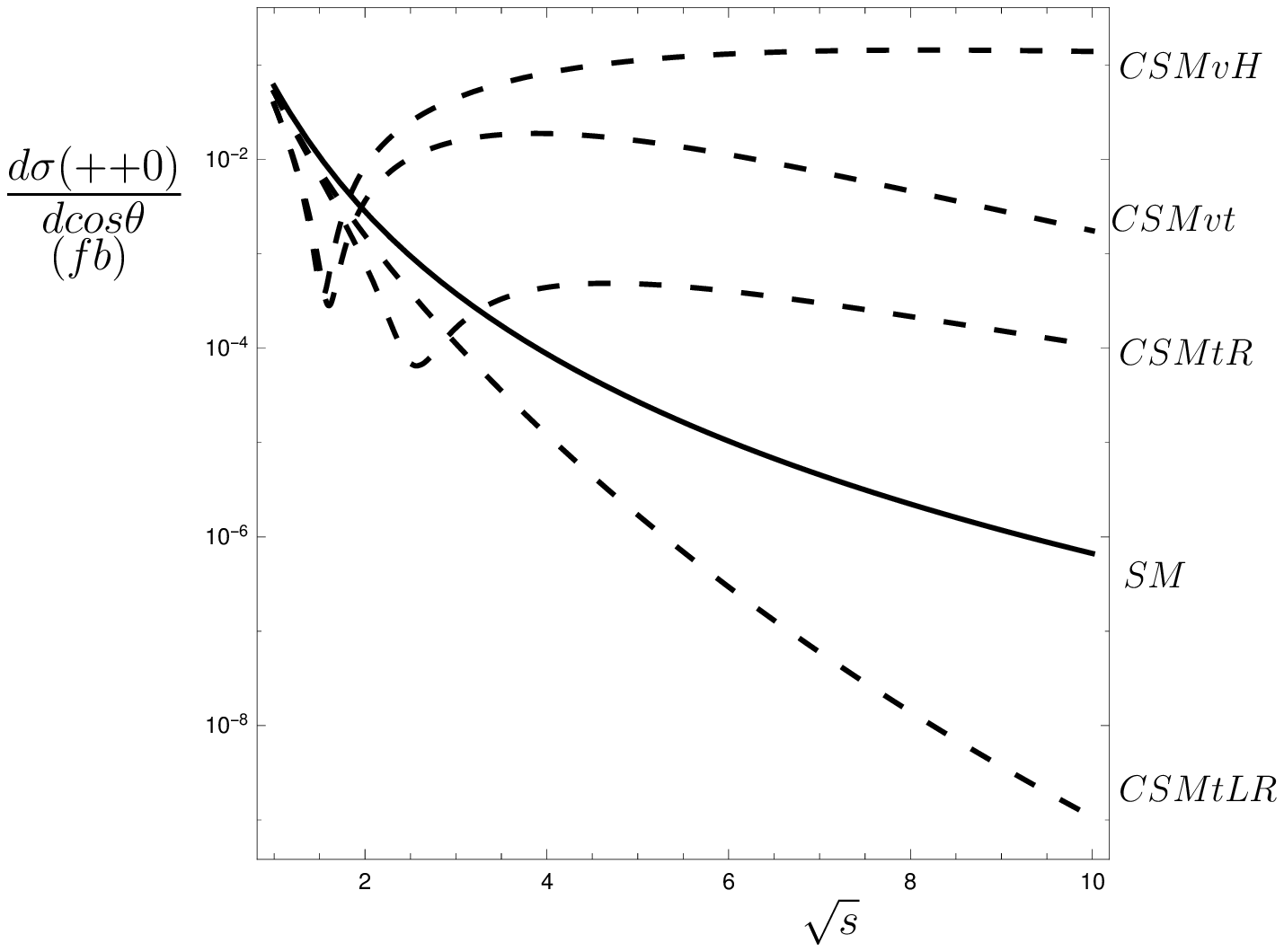, height=6.cm} \hspace{0.2cm}
\epsfig{file=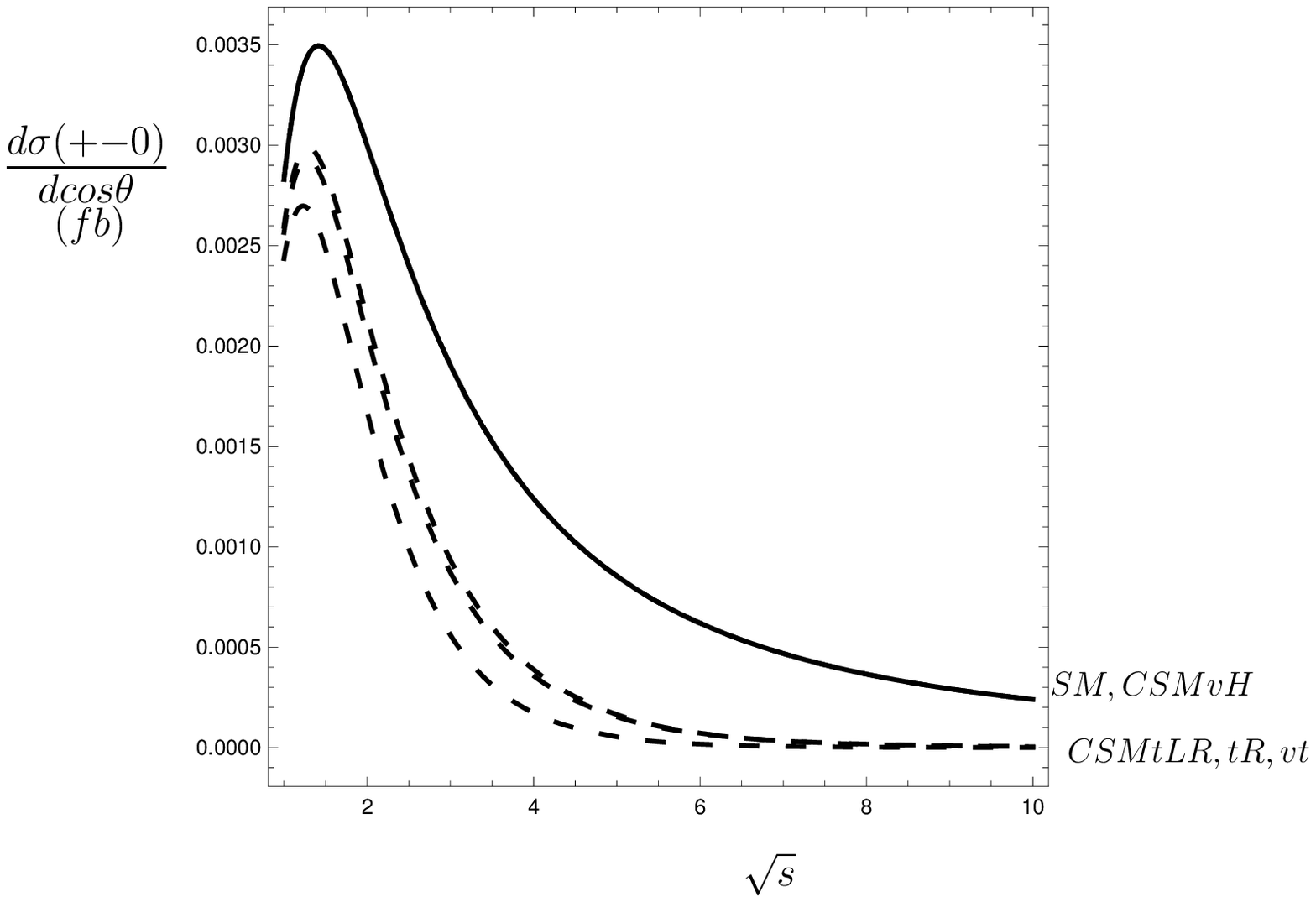, height=6.cm}
\]\\
\[
\epsfig{file=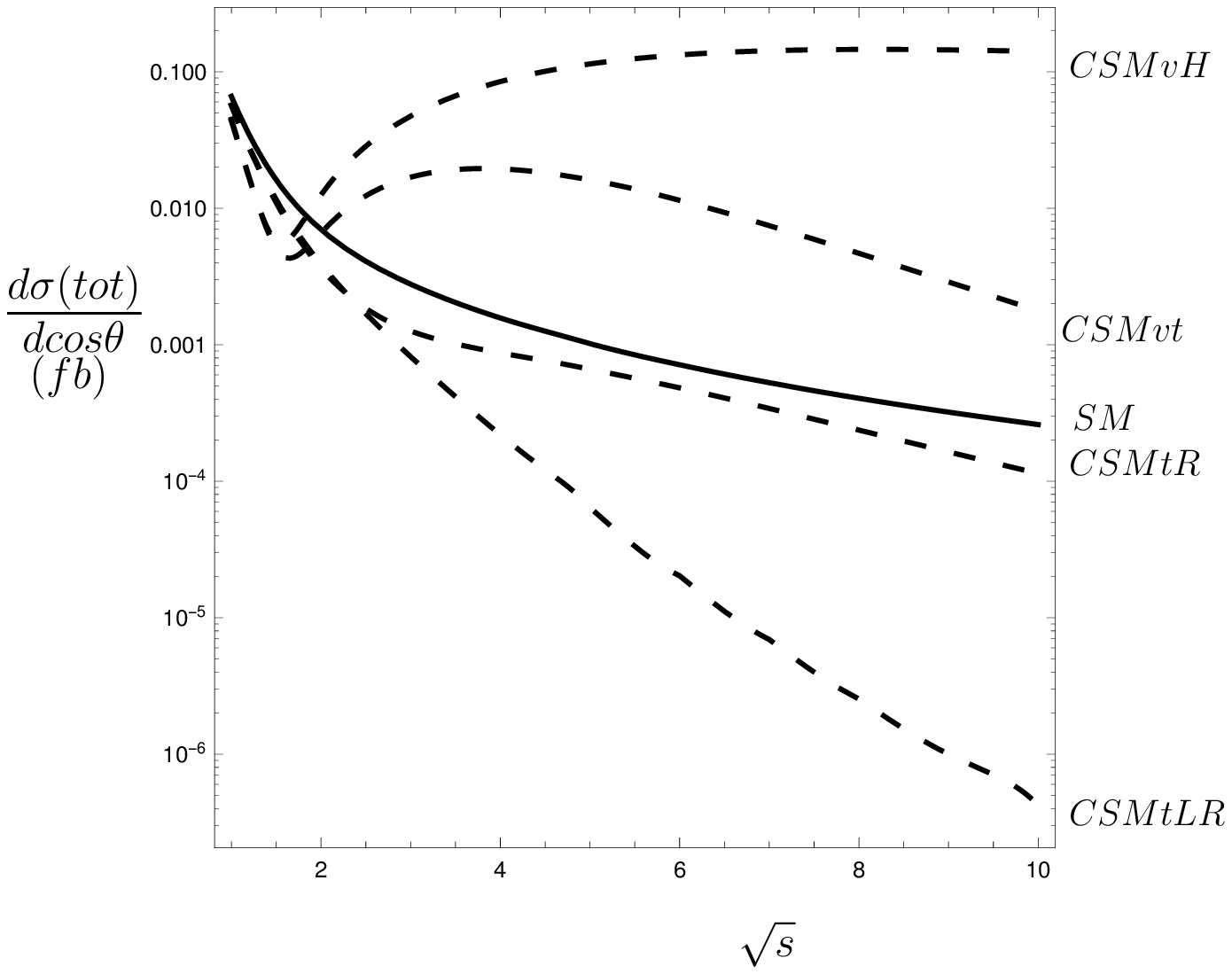, height=8.cm}
\]\\
\vspace{-1cm}
\caption[1] {Cross sections ($\pm\pm0$),($\pm\mp0$) and tot, in SM, CSMtLR, CSMtR, CSMvt, CSMvH cases.}
\end{figure}

\clearpage
\begin{figure}[p]

\[
\epsfig{file=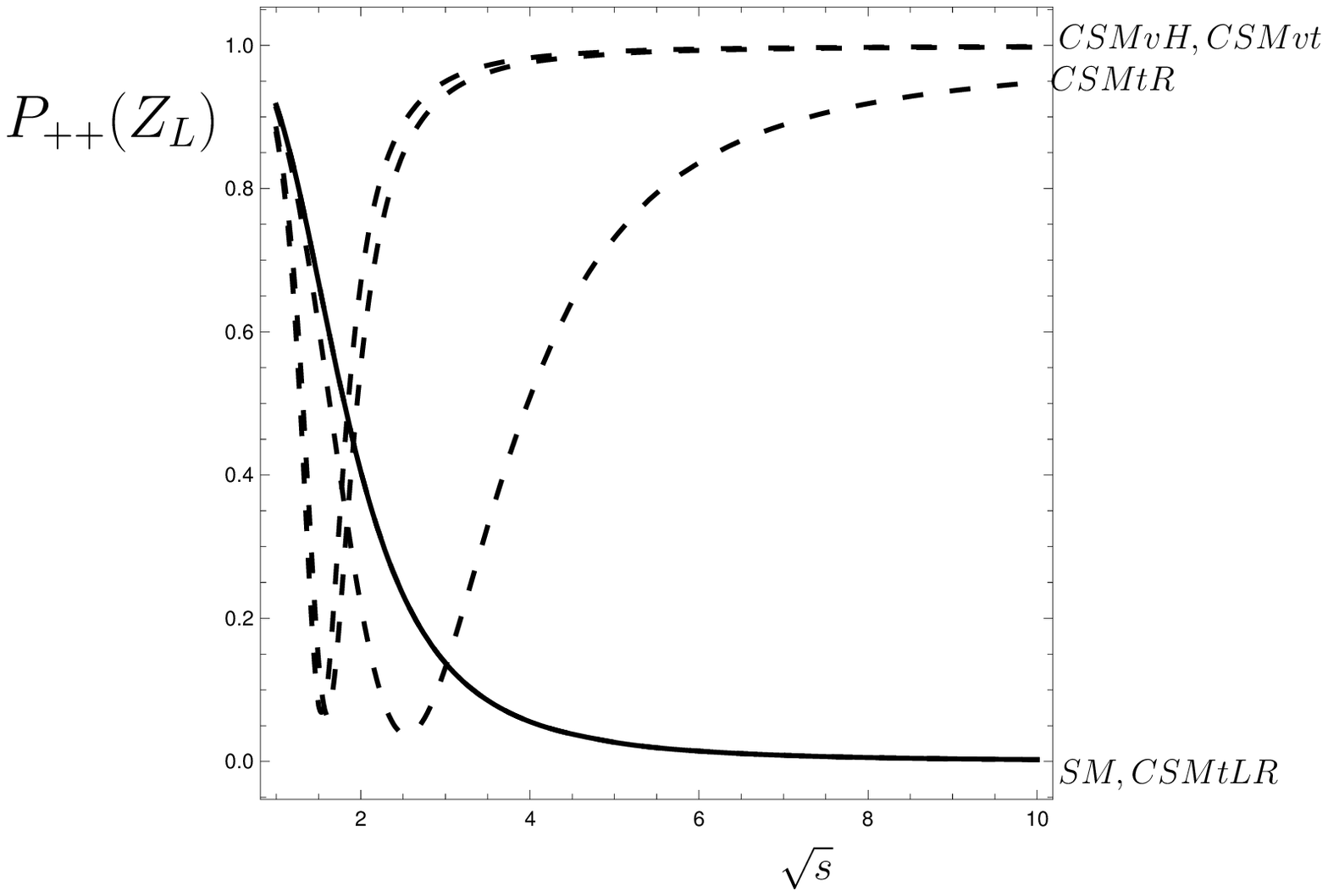, height=6.cm}\hspace{-0.2cm}
\epsfig{file=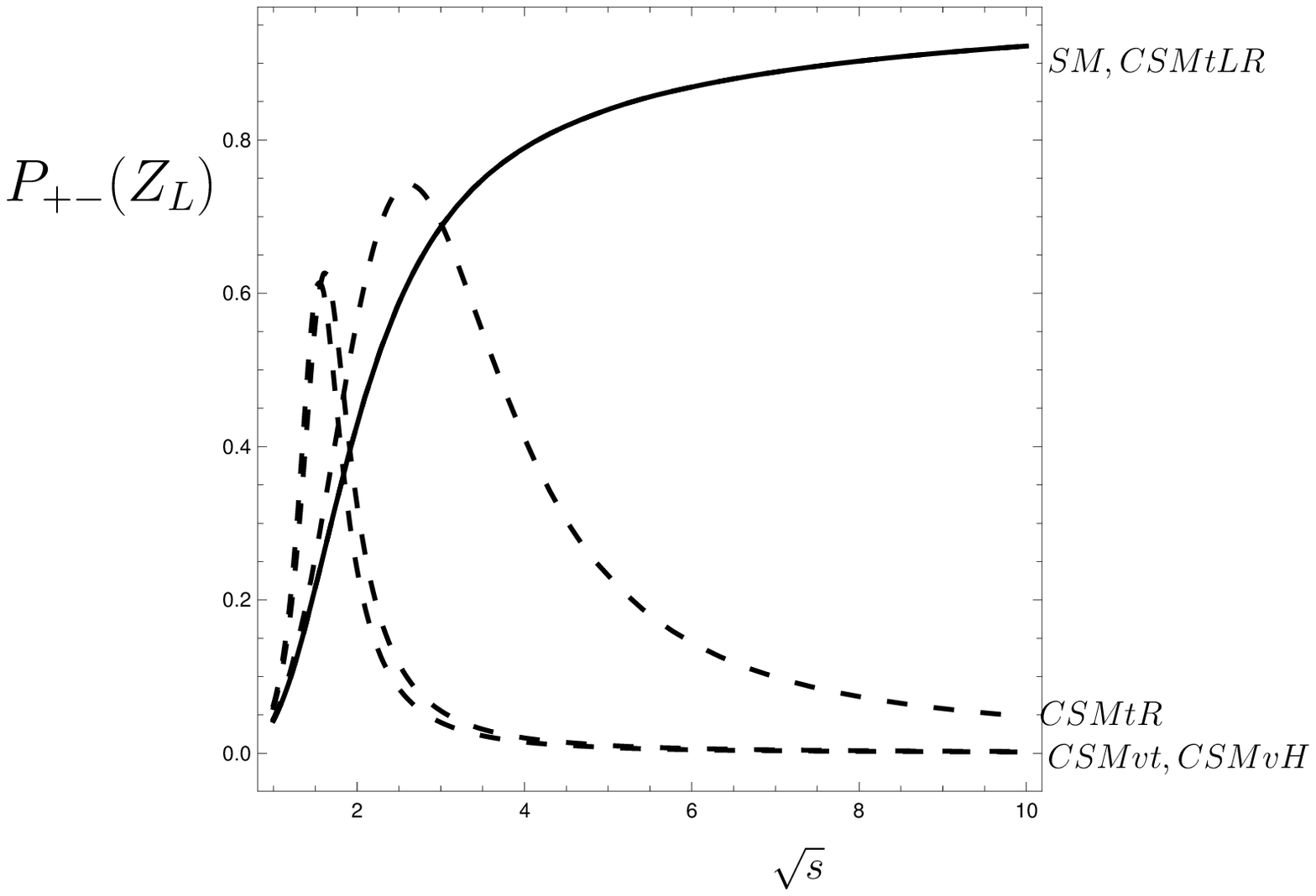, height=6.cm}
\]\\
\vspace{-1cm}
\[
\epsfig{file=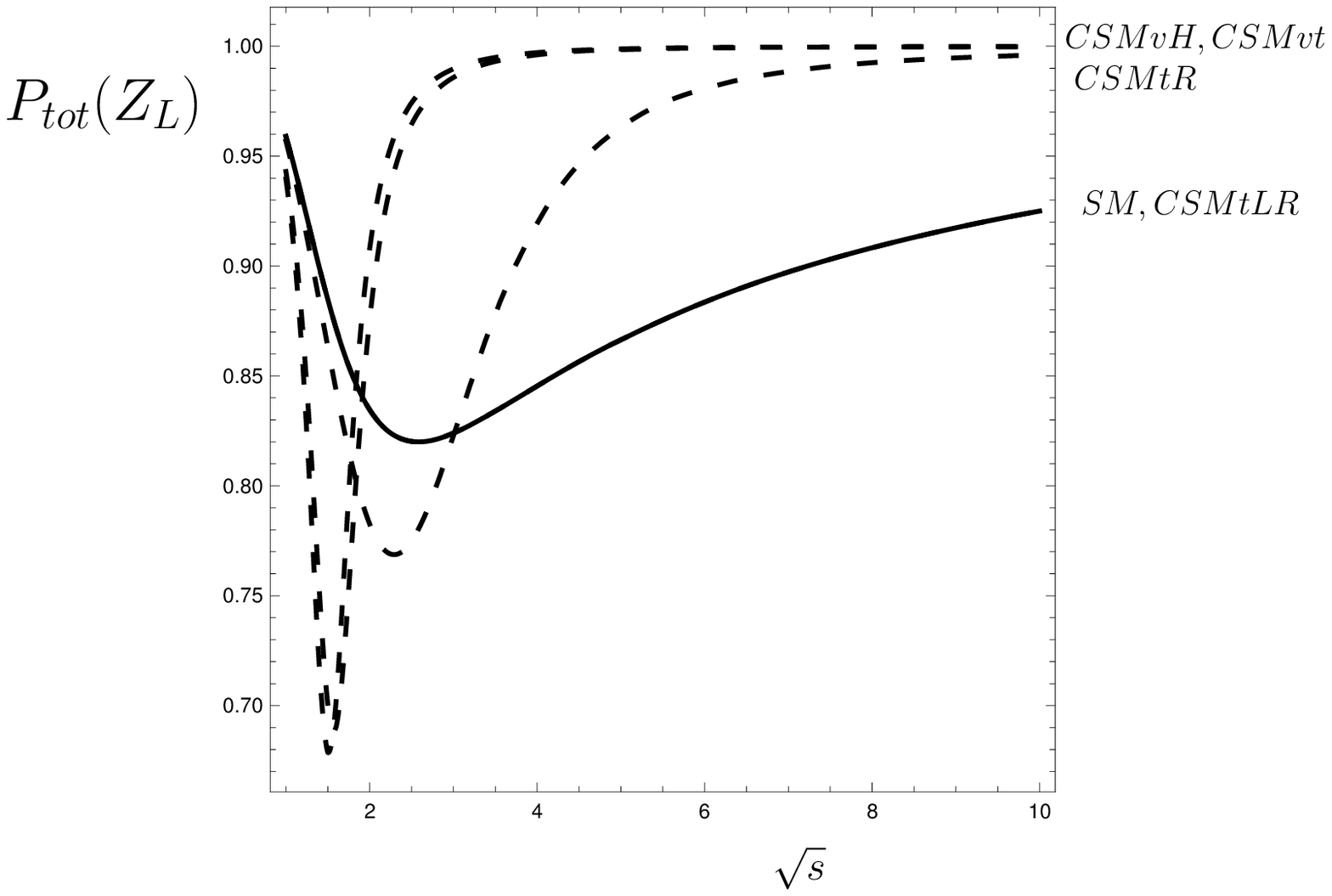, height=8.cm}
\]\\
\caption[1] {$Z_L$ rate from $\pm\pm$,$\pm\mp$ and tot cross sections
in SM, CSMtLR, CSMtR, CSMvt, CSMvH cases.}
\end{figure}

\clearpage
\begin{figure}[p]
\[
\epsfig{file=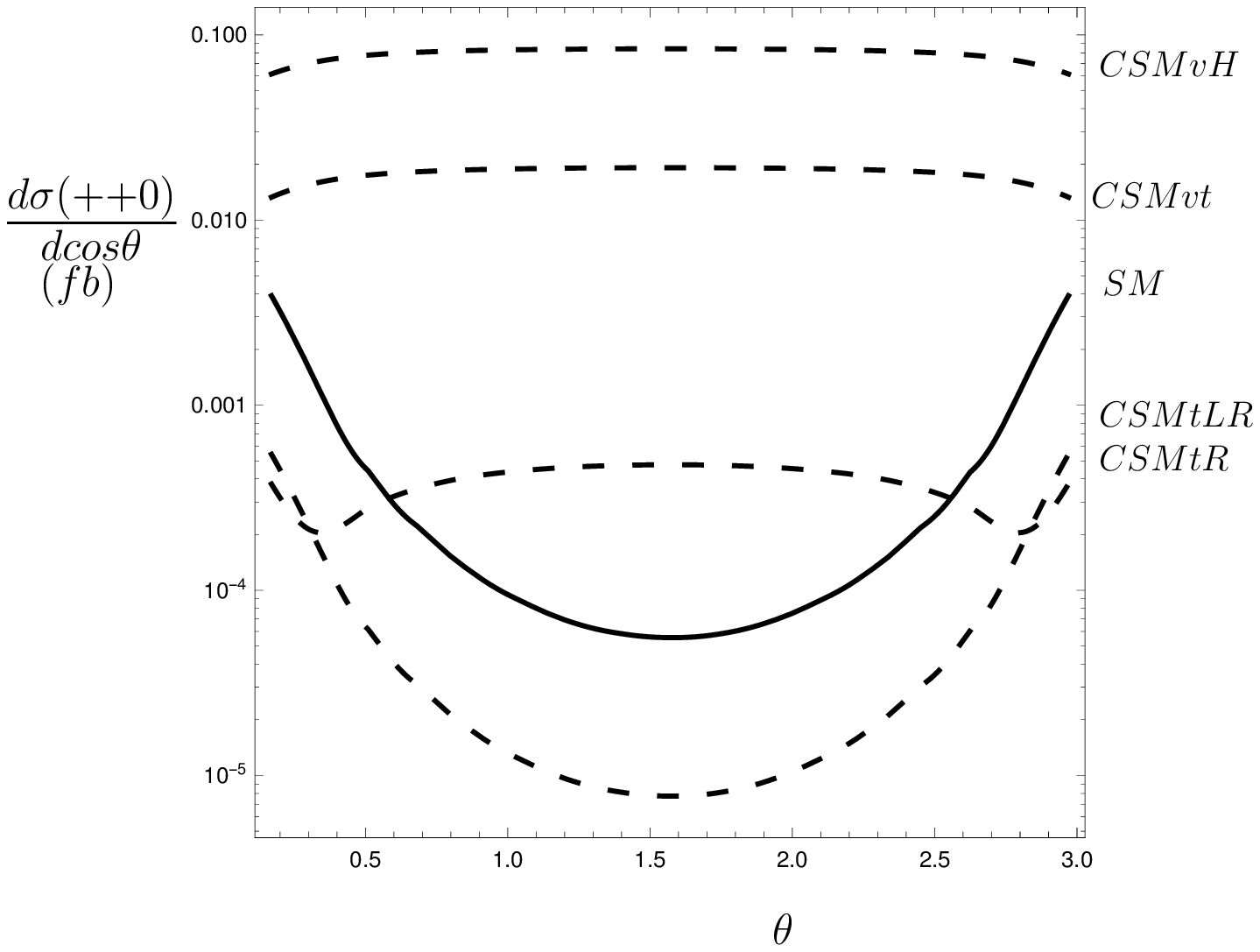, height=6.cm} \hspace{0.1cm}
\epsfig{file=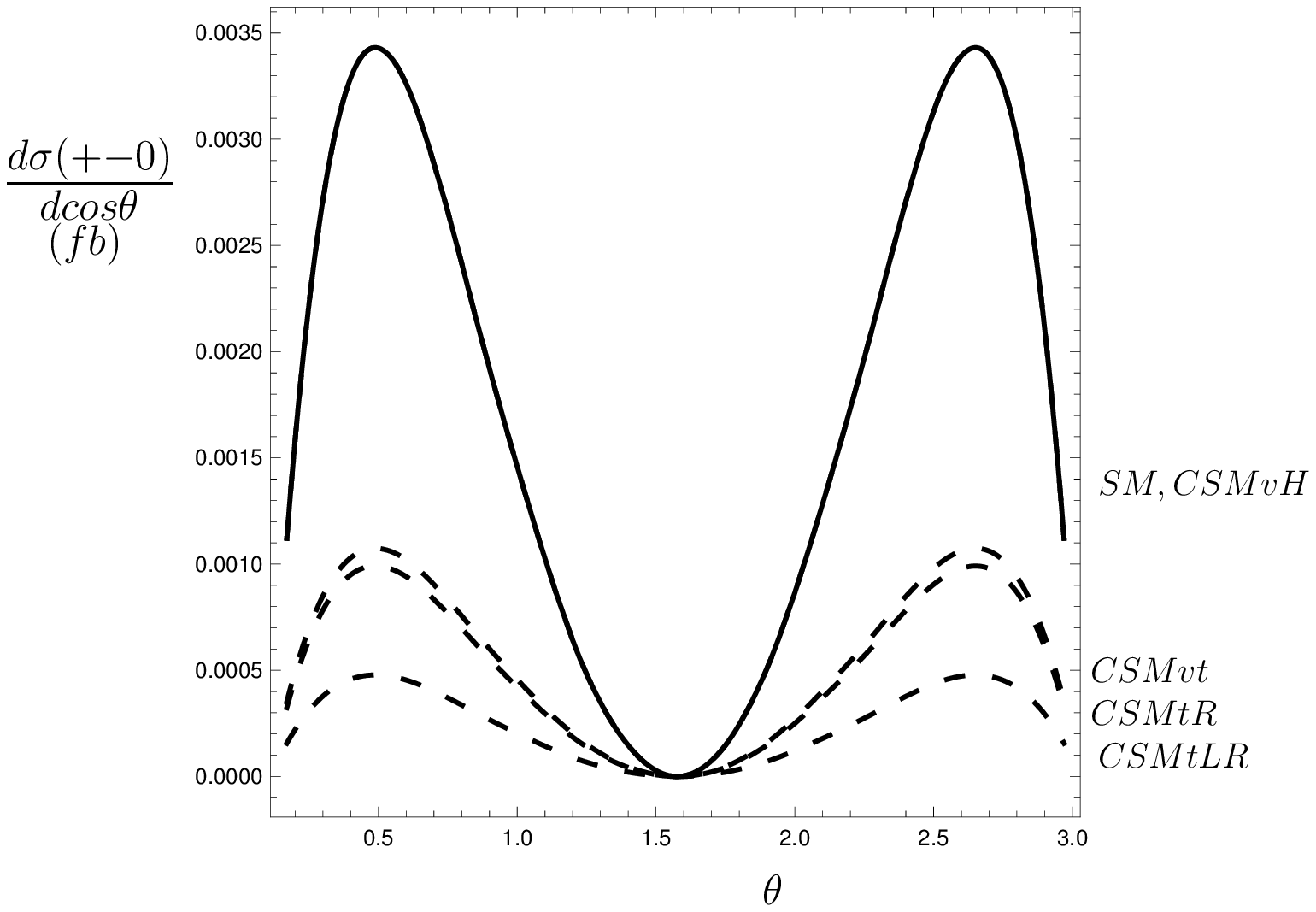, height=6.cm}
\]\\
\[
\epsfig{file=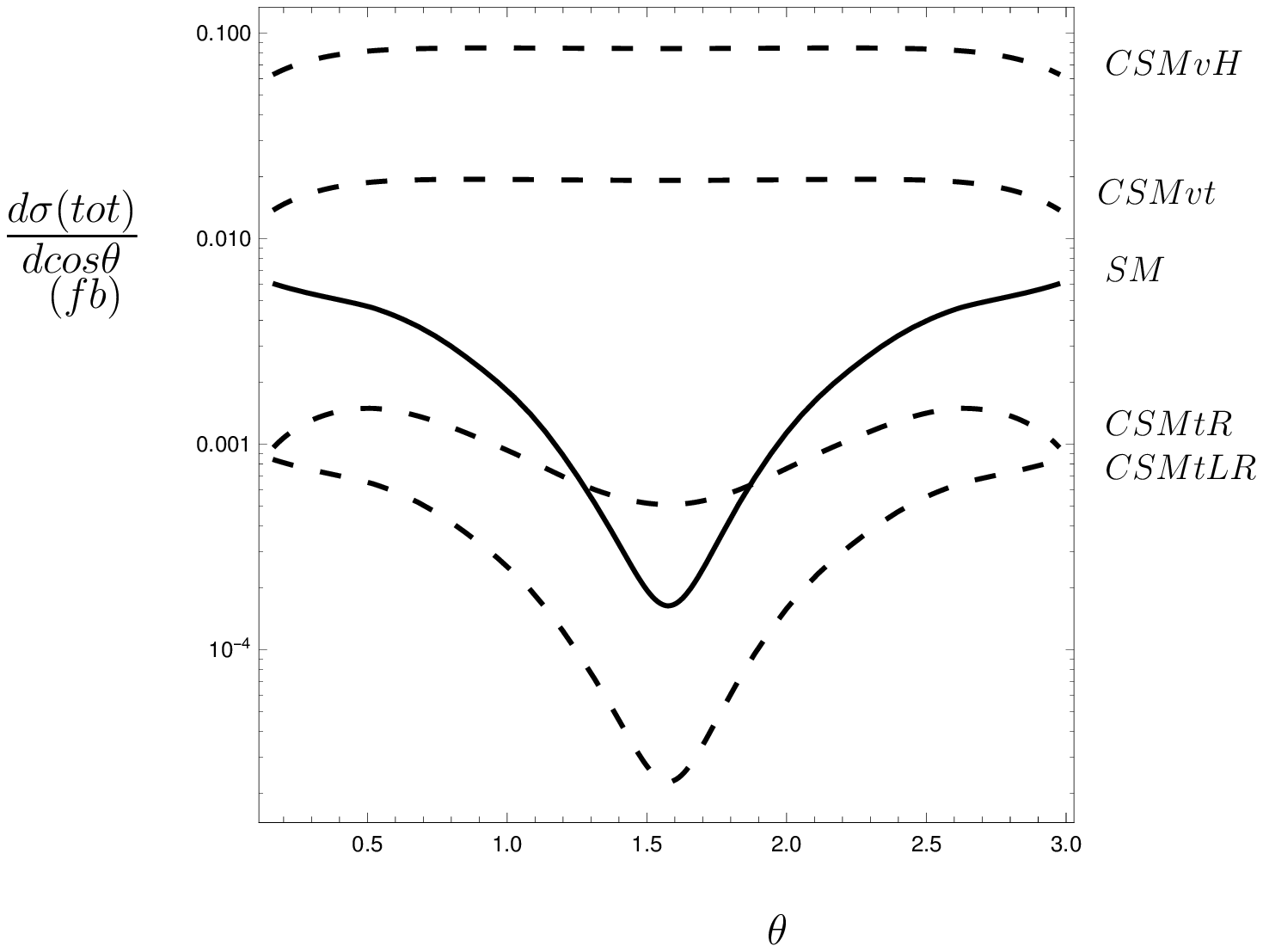, height=8.cm}
\]\\
\vspace{-1cm}
\caption[1] {Angular distribution of the ($\pm\pm0$),($\pm\mp0$) and tot
cross sections,
in SM, CSMtLR, CSMtR, CSMvt, CSMvH cases.}
\end{figure}

\clearpage
\begin{figure}[p]

\[
\epsfig{file=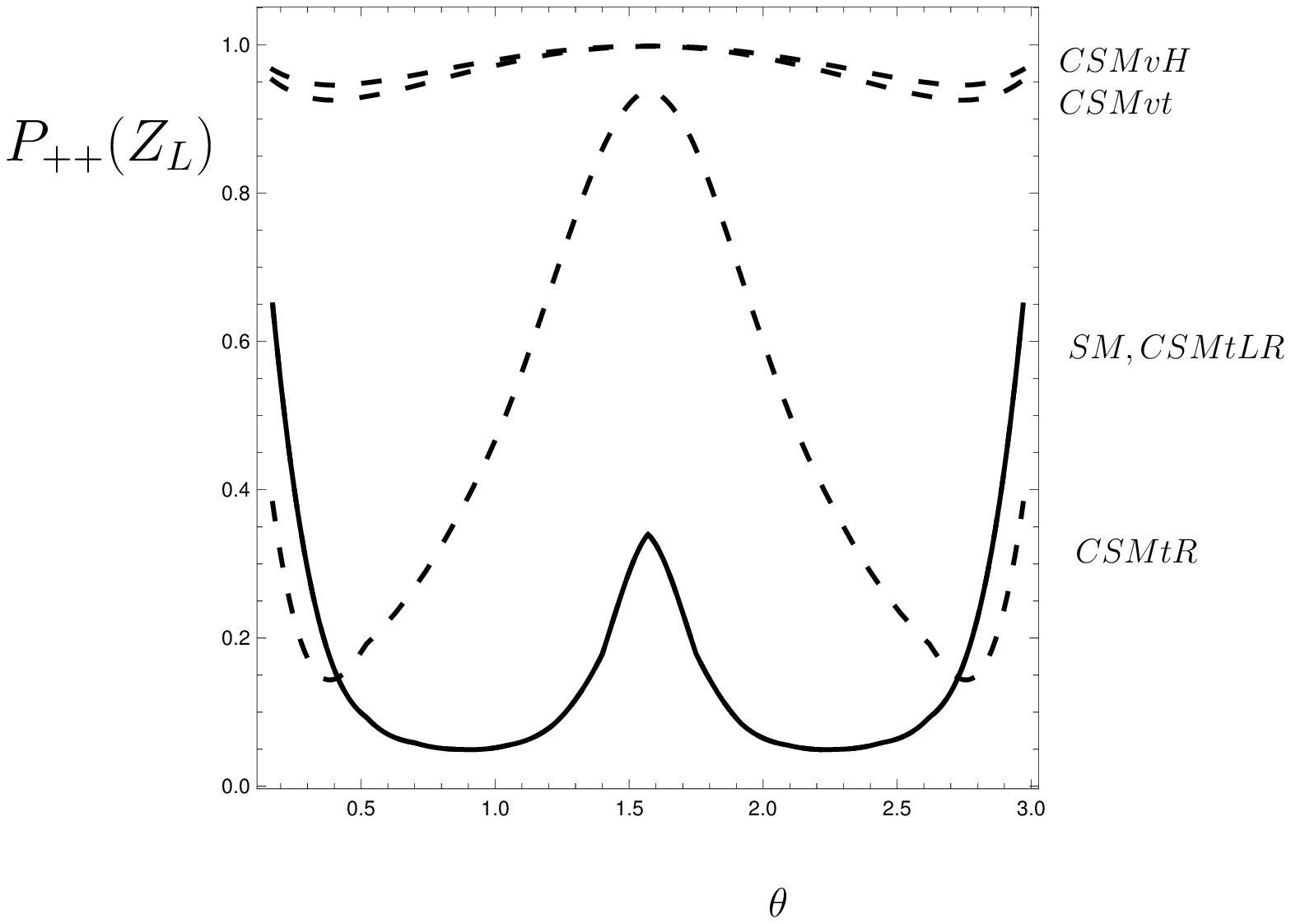, height=6.cm}\hspace{-0.2cm}
\epsfig{file=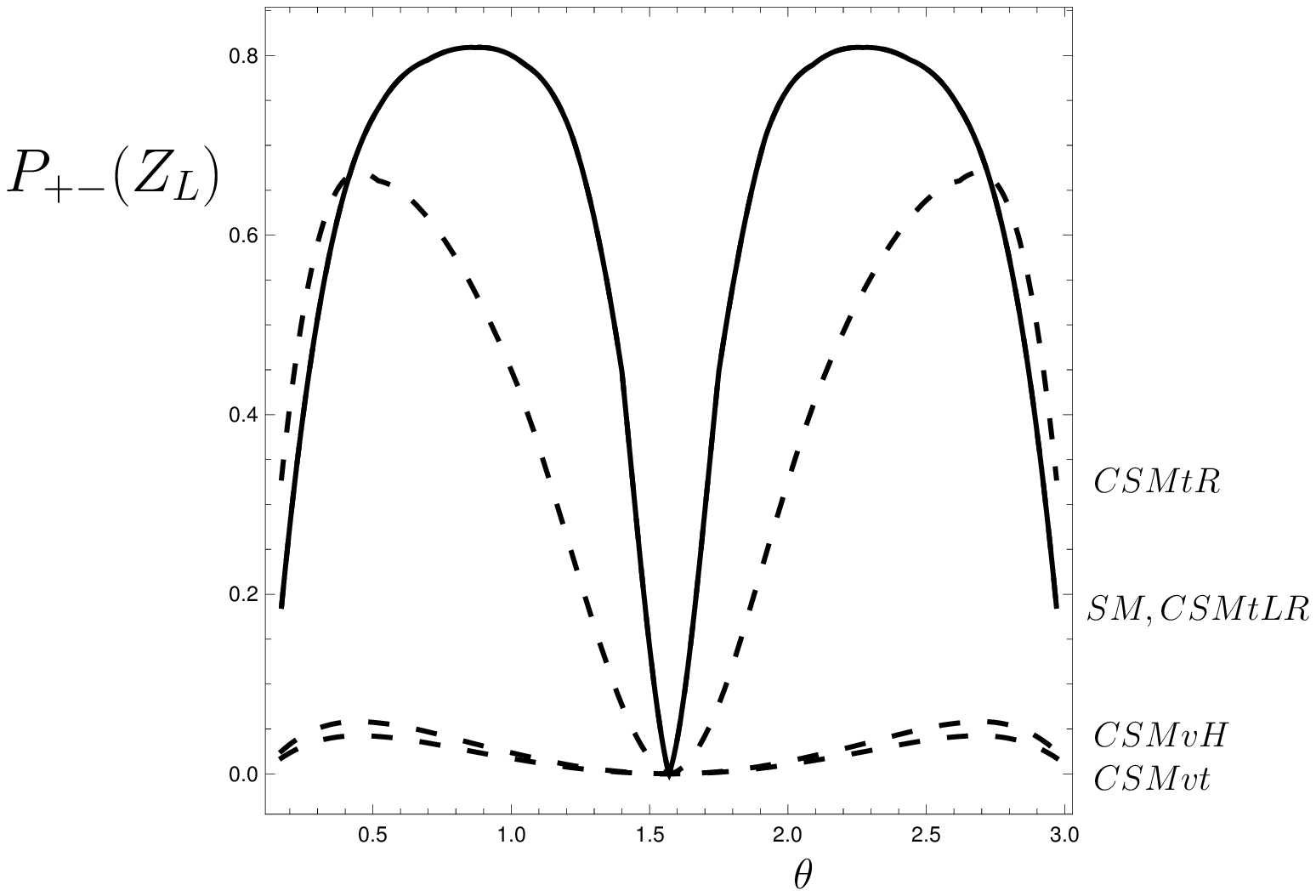, height=6.cm}
\]\\
\vspace{-1cm}
\[
\epsfig{file=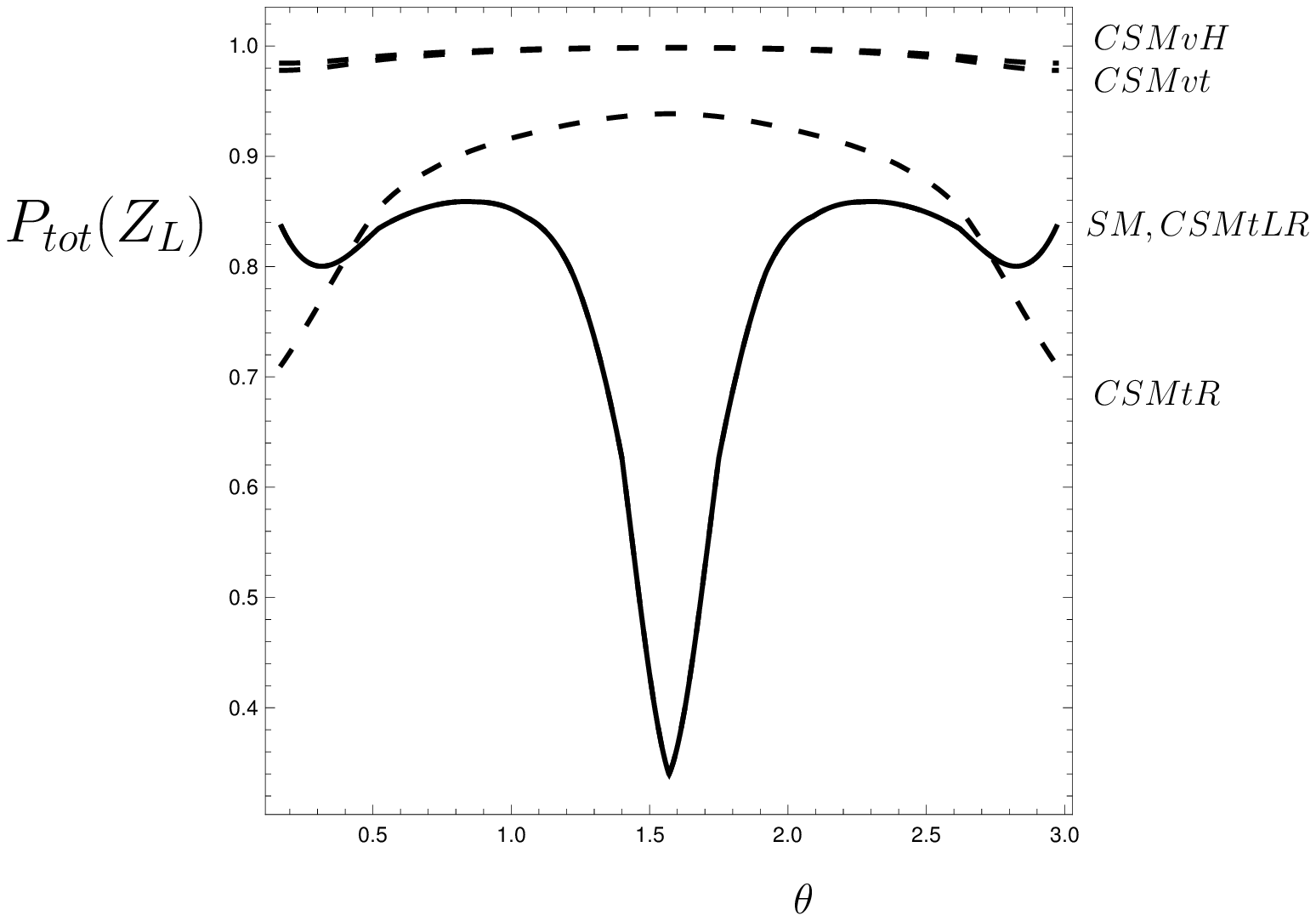, height=8.cm}
\]\\
\caption[1] {Angular distribution of the $Z_L$ rate in SM, CSMtLR, CSMtR, CSMvt, CSMvH cases.}
\end{figure}

\clearpage

\end{document}